\documentclass[aps,prb,amsmath,superscriptaddress,twocolumn,10pt,longbibliography,nofootinbib]{revtex4-2}

\usepackage{graphicx}  
\usepackage{dcolumn}  

\usepackage[T1]{fontenc}  
\usepackage{textcomp}  
\IfFileExists{mtpro2.sty}{
  \usepackage[subscriptcorrection, nofontinfo, mtphrb, mtpcal, mtpfrak]{mtpro2}  
}{
  \usepackage{amssymb}
  \usepackage{amsfonts}
}

\usepackage{siunitx}  
\usepackage{enumitem}  
\usepackage{epstopdf}  
\usepackage{hyperref}  
\usepackage[caption=false,subrefformat=parens,labelformat=parens,justification=RaggedRight,singlelinecheck=true]{subfig}
\usepackage{amsthm}  
\usepackage{tikz-cd}  
\usepackage[italicdiff]{physics}  
\usepackage[textsize=scriptsize, color=orange!60]{todonotes}
\usepackage{simpler-wick}
\usepackage{tikz}
\usepackage{mathtools}
\usepackage{booktabs}
\usepackage{multirow}
\usepackage{xcolor, soul}
\usepackage[only,llbracket,rrbracket]{stmaryrd}  
\usepackage{tabularray}

\usepackage{natbib}
\usepackage{url}
\usepackage{textcase}

\hypersetup{
colorlinks=true,
linkcolor=blue,
citecolor=blue,
urlcolor=blue
}  

\newcolumntype{C}{>{$}c<{$}}  
\newcolumntype{L}{>{$}l<{$}}  
\newcolumntype{R}{>{$}r<{$}}  
\newcolumntype{.}{D{.}{.}{3.12}}  
\newcolumntype{_}{D{.}{.}{10.5}}  

\makeatletter
\renewcommand*\env@matrix[1][\arraystretch]{%
  \edef\arraystretch{#1}%
  \hskip -\arraycolsep
  \let\@ifnextchar\new@ifnextchar
  \array{*\c@MaxMatrixCols c}}
\makeatother  

\ExplSyntaxOn

\cs_new_protected:Nn \ben_matrix_preamble:nn
 {
  \seq_set_split:Nnn \l_tmpa_seq { | } { #2 }
  \seq_set_map:NNn \l_tmpb_seq \l_tmpa_seq { *{##1}{#1} }
  \exp_args:Ne \array{ @{} \seq_use:Nn \l_tmpb_seq { | } @{} }
 }

\clist_map_inline:nn { {},b,B,p,v,V }
 {
  \use:e
   {
    \NewDocumentEnvironment{#1matrix+}{O{c}m}
     {
      \left\str_case:nnF { #1 } {{b}{[}{B}{\{}{p}{(}{v}{|}{V}{\|}}{.}
      \ben_matrix_preamble:nn { ##1 } { ##2 }
     }
     {
      \exp_not:N \endarray
      \right\str_case:nnF { #1 } {{b}{]}{B}{\}}{p}{)}{v}{|}{V}{\|}}{.}
     }
   }
 }

\ExplSyntaxOff

\newtheorem*{theorem*}{Theorem}  
\newtheorem*{definition*}{Definition}  

\usetikzlibrary{arrows, calc, backgrounds}
\makeatletter
\newcommand*{\length}[1]{%
    \@tempcnta\z@
    \@for\@tempa:=#1\do{\advance\@tempcnta\@ne}%
    \the\@tempcnta%
}
\makeatother

\allowdisplaybreaks[1]  
\makeatletter
\@ifclasswith{revtex4-1}{onecolumn}
{\newcommand{\footinsval}{1000}}
{\newcommand{\footinsval}{500}}
\makeatother

\DeclareMathOperator{\Mon}{Mon}

\newcommand{\la}{\langle}  
\newcommand{\ra}{\rangle}  

\newcommand{\iso}{\cong}

\newcommand{\C}{\mathbb{C}}
\newcommand{\Z}{\mathbb{Z}}

\DeclareMathOperator{\vspan}{span}

\DeclareMathOperator{\diag}{diag}
\newcommand{\trans}[1]{{#1}^{\top}}

\DeclarePairedDelimiter\bigspinor{[}{]}

 \newcommand{\bsigma}{\boldsymbol{\sigma}}

\newcommand{\colinddummy}{\bullet}
\newcommand{\perm}{p}

\begin{document}

\count\footins=\footinsval

\title{Non-Abelian Combinatorial Gauge Theory}

\author{Hongji Yu}
\affiliation{Physics Department, Boston University, Boston, MA, 02215, USA}
\author{Dmitry Green}
\affiliation{AppliedTQC.com, ResearchPULSE LLC, New York, NY 10065, USA}
\author{Claudio Chamon}
\affiliation{Physics Department, Boston University, Boston, MA, 02215, USA}

\date{\today}

\begin{abstract}
  Building on the principle of combinatorial gauge symmetry, lattice
  gauge theories can be formulated with only one- and two-body
  interactions that ensure the exact realization of the symmetry
  rather than its approximate emergence in a perturbative regime. This
  paper extends the framework to encompass generic non-Abelian finite
  gauge groups by expanding on previous work that developed the theory
  for finite Abelian gauge groups and presented one non-Abelian
  example.
\end{abstract}

\maketitle


\section{Introduction}





The development of realistic models of quantum spin liquids is a
prominent area of research in condensed matter physics, as it paves
the way for experimental realizations of topological order that could
lead to new approaches to quantum
computation.~\cite{doi:10.1142/S0217979290000139,KITAEV20032,doi:10.1063/1.1499754,Ebadi2021}
Recently, the concept of combinatorial gauge symmetry has emerged as a
promising framework for developing physically realizable models of
quantum spin liquids.~\cite{PhysRevLett.125.067203} Notable
advancements include the construction of Abelian \(\Z_2\) and \(\Z_3\)
gauge theories using two-body
Hamiltonians.~\cite{PhysRevLett.125.067203,PRXQuantum.2.030327} In
addition to proposed realizations of these models using
superconducting wire arrays~\cite{PRXQuantum.2.030327,PRXQuantum.2.030327,PRXQuantum.2.030341},
one such model has been experimentally studied on a programmable
quantum annealing device~\cite{PhysRevB.104.L081107}, thus validating
the underlying principles of the method.  A comprehensive framework
for constructing this sort of models with Abelian gauge groups was
shown in \citet{10.21468/SciPostPhysCore.7.1.014}, establishing the
versatility of this approach.  Furthermore, a model with quaternion
combinatorial gauge symmetry has been constructed~\cite{Green2023},
showing that it is possible to leap from Abelian to non-Abelian gauge
theories.

This paper aims to elucidate the general process for constructing
models with arbitrary finite gauge groups, extending the combinatorial
gauge symmetry framework to encompass non-Abelian gauge theories.  Our
approach leverages the same mathematical foundations while introducing
a novel method for embedding gauge degrees of freedom into spin
systems, thereby making it more realistic to realize these models in
quantum computing devices.


The outline of this paper is as follows.  First we present the
construction for the simplest non-Abelian group \(S_3\) to illustrate
the full process of building a model with non-Abelian combinatorial
gauge symmetry.  Next, we describe the general theory of combinatorial
gauge symmetry for arbitrary finite gauge groups.  This includes the
method of embedding gauge degrees of freedom into systems of physical
spins, as well as the generalized procedure of introducing auxiliary spins and coupling them to the
gauge variables via a specially designed two-body Hamiltonian.

\section{A model with \texorpdfstring{\(S_3\)}{S3} combinatorial gauge symmetry}

\begin{figure}[tb]
  \centering
  \includegraphics[width=0.9\columnwidth]{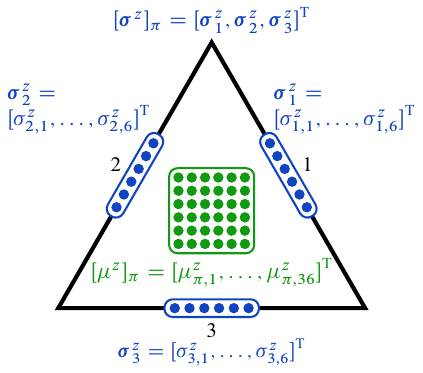}
  \caption{Arrangement of degrees of freedom in an \(S_3\) non-Abelian
    combinatorial gauge theory. On each link of the lattice, there are
    \(6\) spin-\(1/2\) degrees of freedom (blue dots) to represent
    the gauge variables. Around each plaquette, they are coupled via
    an Ising-type interaction to \(36\) spin-\(1/2\) degrees of
    freedom (green dots) located in the center of each plaquette. This
    diagram represents the degrees of freedom in the plaquette with
    index \(\pi\).}
  \label{fig:s3-spin-configuration}
\end{figure}


In this section, we illustrate the process of constructing
combinatorial gauge theories using a model defined on a triangular
lattice that possesses an \(S_3\) gauge symmetry.  As the first step,
we describe a representation of \(S_3\) gauge degrees of freedom in
terms of a \(6\)-spinor.  Next we study the algebraic structure of
gauge transformations that are relevant to the construction.  We then
introduce additional (matter) spins and couple them to the gauge
spinors via a two-body interaction that has the required gauge
symmetries.  Finally, we need to include an additional term to ensure
that the gauge flux is zero in the ground state of our model.

Consider a model with \(6\) spin-\(1/2\) degrees of
freedom \(\sigma_{\lambda,1}^z, \dots, \sigma_{\lambda, 6}^z\) on each
link \(\lambda\) of a triangular lattice. Organize them into a
\(6\)-spinor \(\bsigma^z_\lambda\) --- it will be used to represent one gauge degree of freedom taking values in \(S_3\). 
It is also useful to collect the spinors on the three links
of a plaquette \(\pi\) of the lattice, and concatenate these
three \(6\)-spinors \({\bsigma}^z_1, {\bsigma}^z_2, {\bsigma}^z_3\) into an
\(18\)-spinor
\([\bsigma^z]_\pi = \trans{[{\bsigma}^z_1, {\bsigma}^z_2,
  {\bsigma}^z_3]}\) that will simplify our writing of some of the
interactions.\footnote{The bold letter in the
notation for the gauge spinor \(\bsigma^z_\lambda\) indicates that
this is a multi-component object, in contrast with its components
\(\sigma^z_{\lambda, i}\) and matter spins \(\mu^z_{\pi,a}\), defined later. On the other hand, when a symbol is surrounded by brackets and subscripted with \(\pi\), it stands for a spinor composed of all objects of one type associated with the plaquette \(\pi\). Thus \([\bsigma^z]_\pi\) is a spinor consisting of all gauge spinors \(\bsigma^z_\lambda\) that surround the plaquette \(\pi\), and similarly \([\mu^z]_\pi\) is a spinor of all matter spins located within the plaquette \(\pi\).} Note that neighboring plaquettes share links and
thus their corresponding \(6\)-spinors.

In every plaquette \(\pi\) of the lattice we place \(36\) additional spin-\(1/2\)
degrees of freedom \(\mu^z_{\pi, 1}, \dots, \mu^z_{\pi, 36}\), which
are also grouped into a spinor \([\mu^z]_{\pi}\). 
These ``matter''
degrees of freedom enable the CGS construction, since we will use them to enforce the
gauge symmetry through the coupling between them and the gauge degrees of freedom.
Figure~\ref{fig:s3-spin-configuration} illustrates the arrangement of
these spins.

The Hamiltonian of the system can be written as a sum over all
plaquettes of the lattice:
\begin{align}
  \label{eq:H-as-sum-of-plaquettes}
  H = \sum_\pi \left(
  H^\text{proj}_\pi
  +
  H^\text{CGS}_\pi
  +
  H^\text{flux}_\pi
  \right)
  \;.
\end{align}
Around each plaquette \(\pi\), we have the following three terms, each consisting of only one- and two-spin interactions,
\begin{itemize}[wide]
  \item 
  $H^\text{proj}_\pi$ restricts the link \(6\)-spinors on each of the three links of the plaquette to one of the 6 states associated to the elements of $S_3$ via an Ising interaction between spins on each link.
  We can express $H^\text{proj}_\pi = \frac{1}{2} \sum_{\lambda\in\pi}H^\text{proj}_\lambda$, where $H^\text{proj}_\lambda$ enforces the constraints on each individual link $\lambda$. 
  (The factor of 1/2 accounts for each link being shared by two plaquettes.)

  \item 
  $H^\text{CGS}_\pi$ is also an Ising interaction that couples the gauge and matter spins associated with the plaquette:
  \begin{align}
    H^\text{CGS}_\pi = - J_c\;\;\trans{[\mu^z]_\pi}
    \; W\;
    [{\bsigma}^z]_\pi
    \;,
  \end{align}
  where the $36\times 18$ matrix of couplings $W$ will be constructed explicitly below. 
  While the other two terms are gauge symmetric due to their permutation symmetry, this term possesses a key property that gives it the desired gauge symmetry.
  Namely, if we permute the \emph{columns} of the coupling matrix \(W\) in ways determined algebraically by the gauge symmetries, we can undo the transformation by another permutation of its \emph{rows}.\footnote{This property makes these coupling matrices analogous to certain \emph{combinatorial designs} such as Hadamard matrices. This is the origin of the term \emph{combinatorial gauge symmetry}.}  
  As a result, our Hamiltonian is invariant under joint transformations of the gauge spinors \({\bsigma}^z_\lambda\) on the links and matter spins \([\mu^z]_\pi\) on the centers of the plaquettes, and the action on the gauge part of the system mirrors the behavior of a pure gauge theory defined on the same lattice.

  \item
  $H^\text{flux}_\pi$ contains a uniform longitudinal field on the matter spins to select the zero flux sector as the ground state.

\end{itemize}

Below we construct of these three types of one- and two-body interaction terms explicitly. 
Because the Hamiltonian~\eqref{eq:H-as-sum-of-plaquettes} is a sum of identical
terms for each plaquette $\pi$, and we can focus on one plaquette at a
time, for simplicity we will suppress the subscript \(\pi\) in the
subsections that follows, so that \([\bsigma^z] \equiv [\bsigma^z]_\pi\) and \([\mu^z]\equiv [\mu^z]_\pi\) should be understood.
The full lattice Hamiltonian can be recovered by attaching the plaquette label \(\pi\) to the spinors \([\bsigma^z]\) and \([\mu^z]\) and summing over \(\pi\).

\subsection{Spin representation of \texorpdfstring{\(S_3\)}{S3} gauge variables}


We adopt an ordering of the elements of \(S_3\) as follows,
\begin{equation}
\begin{split}
S_3 & = \{1, c, c^2, s, sc, sc^2\} \\
& =
\left\{\left(\begin{smallmatrix}
  1 & 2 & 3 \\
  1 & 2 & 3
\end{smallmatrix}\right), 
\left(\begin{smallmatrix}
  1 & 2 & 3 \\
  2 & 3 & 1
\end{smallmatrix}\right), 
\left(\begin{smallmatrix}
  1 & 2 & 3 \\
  3 & 1 & 2
\end{smallmatrix}\right), \right.\\
& \qquad\left.
\left(\begin{smallmatrix}
  1 & 2 & 3 \\
  2 & 1 & 3
\end{smallmatrix}\right), 
\left(\begin{smallmatrix}
  1 & 2 & 3 \\
  3 & 2 & 1
\end{smallmatrix}\right), 
\left(\begin{smallmatrix}
  1 & 2 & 3 \\
  1 & 3 & 2
\end{smallmatrix}\right)
\right\}\ ,
\label{eq:s3-elements}
\end{split}
\end{equation}
and we choose the cyclic permutation
\(c = \left(\begin{smallmatrix}
  1 & 2 & 3 \\ 
  2 & 3 & 1
\end{smallmatrix}\right)\) 
and transposition
\(s = \left(\begin{smallmatrix}
  1 & 2 & 3 \\
  2 & 1 & 3
\end{smallmatrix}\right)\)
as the generators, so that the group has the presentation \(S_3 = \la c, s \mid c^3 = s^2 = (cs)^2 = 1 \ra\).

To represent an \(S_3\) gauge variable, we place six spins on each link \(\lambda\), which forms a \(6\)-spinor
\begin{equation}
  \bsigma^z_\lambda =
  \begin{bmatrix}
    \sigma^z_{\lambda, 1} \\
    \sigma^z_{\lambda, 2} \\
    \vdots \\
    \sigma^z_{\lambda, 6}
  \end{bmatrix}\ .
\end{equation}
We project the spinor into the six-dimensional subspace where \(\sum_{i = 1}^6 \sigma_{\lambda, i}^z = 4\), using the following interaction,
\begin{equation}
  H_{\lambda}^{\text{proj}}= J_p \left(\sum_{i = 1}^{6} \sigma_{\lambda, i}^z - 4\right)^2\ ,
  \label{eq:s3-hamiltonian-projection-term}
\end{equation}
where the coefficient \(J_p\) will be sufficiently large so that all states that do not represent the gauge variable are lifted to energy levels much larger than the gap of the CGS coupling term \(H^\text{CGS}\).
The gauge degree of freedom lives in this subspace, which we denote \(\mathcal{H}^G_\lambda\).
Assuming an ordering of the group elements of the gauge group \(S_3\), there is a canonical basis for the gauge degree of freedom, \(\{\ket{g_i}: g_i \in G\}\), labeled by the group elements \(g_i\), which we map to the spinor states, so that the \(i\)-th basis state corresponds to the spinor state whose \(i\)-th component takes value \(-1\).
Symbolically, this definition is expressed by the condition \(\sigma^z_{\lambda, j} \ket{g_i} = (1 - 2 \delta_{ij}) \ket{g_i}\).
Thus the spinor operator \(\bsigma^z_\lambda\) takes the following values \(v(g_i)\) on the state \(\ket{g_i}\),
\begin{equation}
\begin{aligned}
  v(1) & = \trans{[- + + + + +]}\ , & v(s) & = \trans{[+ + + - + + ]}\ ,\\
  v(c) & = \trans{[+ - + + + +]}\ , & v(s c) & = \trans{[+ + + + - +]} \ ,\\
  v(c^2) & = \trans{[+ + - + + +]}\ , & v(s c^2) & = \trans{[+ + + + + -]}\ .
\end{aligned}
\label{eq:s3-state-labels}
\end{equation}
We will refer to these vectors collectively as \emph{state labels}.

Next, define a set of permutation matrices \(\ell(h)\) and \(r(h)\) which act on the vectors \(v(g)\) by group multiplication on the left and on the right, respectively
\begin{equation}
\begin{aligned}
  \ell(h) v(g) & = v(h g)\ , \\
  r(h) v(g) & = v(g h^{-1})\ .
\end{aligned}
\label{eq:v-l-r-relations}
\end{equation}
Since the left- and right-multiplication maps \(g\mapsto hg\) and \(g \mapsto gh^{-1}\) permute the group elements, the above relations imply that \(\ell(h)\) and \(r(h)\) are permutation matrices.
This is a key property of \(\ell(h)\) and \(r(h)\), because as we shall see, these matrices serve the dual purpose of describing the effects of \(S_3\)-gauge transformations on the gauge spinors, as well as corresponding to physical symmetries of the component spins that make up the spinor.
Importantly, permutations preserve the commutators of the spin operators \(\sigma^z_{\lambda, i}\).
The matrices that represent left- and right-multiplication by the generators of the group are
\begin{align}
  \ell(c) & = \begin{bmatrix}[0.9]
    0 & 0 & 1 & 0 & 0 & 0 \\
    1 & 0 & 0 & 0 & 0 & 0 \\
    0 & 1 & 0 & 0 & 0 & 0 \\
    0 & 0 & 0 & 0 & 1 & 0 \\
    0 & 0 & 0 & 0 & 0 & 1 \\
    0 & 0 & 0 & 1 & 0 & 0
  \end{bmatrix}\ , & 
  \ell(s) & = \begin{bmatrix}[0.9]
    0 & 0 & 0 & 1 & 0 & 0 \\
    0 & 0 & 0 & 0 & 1 & 0 \\
    0 & 0 & 0 & 0 & 0 & 1 \\
    1 & 0 & 0 & 0 & 0 & 0 \\
    0 & 1 & 0 & 0 & 0 & 0 \\
    0 & 0 & 1 & 0 & 0 & 0
  \end{bmatrix}\label{eq:s3-l-operator-generators}\\
  \intertext{and}
  r(c) & = \begin{bmatrix}[0.9]
    0 & 1 & 0 & 0 & 0 & 0 \\
    0 & 0 & 1 & 0 & 0 & 0 \\
    1 & 0 & 0 & 0 & 0 & 0 \\
    0 & 0 & 0 & 0 & 1 & 0 \\
    0 & 0 & 0 & 0 & 0 & 1 \\
    0 & 0 & 0 & 1 & 0 & 0
  \end{bmatrix}\ , & 
  r(s) & = \begin{bmatrix}[0.9]
    0 & 0 & 0 & 1 & 0 & 0 \\
    0 & 0 & 0 & 0 & 0 & 1 \\
    0 & 0 & 0 & 0 & 1 & 0 \\
    1 & 0 & 0 & 0 & 0 & 0 \\
    0 & 0 & 1 & 0 & 0 & 0 \\
    0 & 1 & 0 & 0 & 0 & 0
  \end{bmatrix}\label{eq:s3-r-operator-generators}
\end{align}
The representations of the other group elements can be computed using the group multiplication law \(\ell(g)\ell(h) = \ell(gh)\) and \(r(g)r(h) = r(gh)\).

\subsection{\texorpdfstring{\(S_3\)}{S3} gauge symmetries on a triangular lattice}
\label{sec:s3-gauge-symmetry}

\begin{figure}[tb]
  \subfloat[A gauge degree of freedom is located on each link, which has an orientation indicated by the arrow. Blue shades highlight the three gauge degrees of freedom, \(g_1, g_2, g_3\), associated with a plaquette on the triangular lattice. They multiply to the flux \(g_1 g_2 g_3\) through the plaquette. ]{
    \label{fig:s3-triangular-lattice-plaquette}
    \includegraphics[width=0.8\columnwidth]{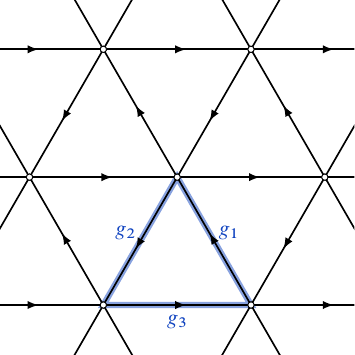}
  }\qquad
  \subfloat[Red arrows show the effect of a gauge transformation generated by a charge \(h\) on one vertex. When an arrow agrees with the orientation of a link, the gauge variable on that link is left-multiplied by \(h\). If the orientations disagree, the gauge variable is right-multiplied by \(h^{-1}\). When a charge is located on one vertex of a plaquette, two links of the plaquette are affected by the gauge transformation, one left-multiplied, one right-multiplied.]{
    \label{fig:s3-triangular-lattice-plaquette-star}
    \includegraphics[width=0.8\columnwidth]{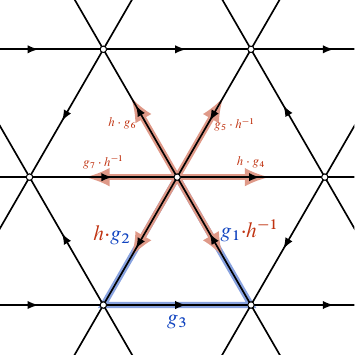}
  }
  \caption{\(S_3\) gauge symmetries on a triangular lattice}
  \label{fig:s3-triangular-lattice}  
\end{figure}

\begin{figure}
  \subfloat[A single charge \(h\) located on one of the vertices of a triangular plaquette induces gauge transformations on two of the adjacent links. On the link that is oriented towards this vertex, the gauge transformation acts as a right multiplication by \(h\), and on the link that is oriented away from the vertex, the transformation is a left multiplication. This gauge transformation can be rewritten as a matrix multiplication on a vector \(\trans{[g_1, g_2, g_3]}\). ]{
    \label{fig:s3-plaquette-gauge-transformation-single}
    \includegraphics[width=0.75\columnwidth]{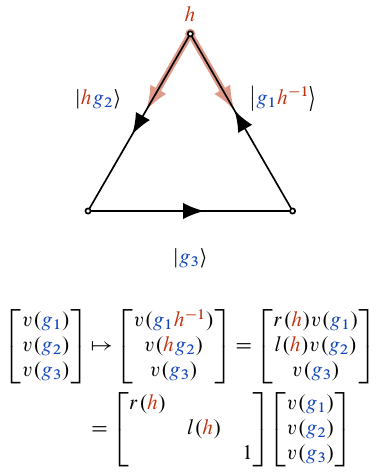}
  }\qquad
  \subfloat[The effect of a general gauge transformation generated by three different \(S_3\) charges, \(h, k, f\), on the gauge variables around a plaquette. Each gauge variable is left- and right-multiplied by different charges. Overall, the product \(g_1 g_2 g_3\) is conserved up to a conjugation.]{
    \label{fig:s3-plaquette-gauge-transformation-all}
    \includegraphics[width=0.72\columnwidth]{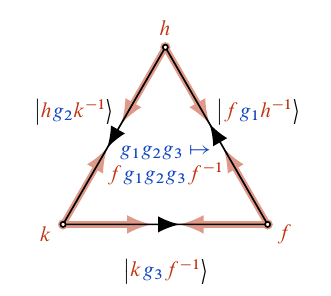}
  }
  \caption{The effect of gauge transformations on the gauge variables around a single plaquette}
  \label{fig:s3-plaquette-gauge-transformation}
\end{figure}

Now we zoom out to one triangular plaquette consisting of three links, and construct an interaction involving the three link variables \(g_1, g_2, g_3\in S_3\) that favors configurations that satisfy the zero-flux condition \(g_1 g_2 g_3 = 1\) 
(see Fig.~\ref{fig:s3-triangular-lattice}\subref{fig:s3-triangular-lattice-plaquette}).
This interaction must be invariant under all gauge transformations that preserve the condition \(g_1 g_2 g_3 = 1\), which are generated by \(S_3\) charges on the vertices adjacent to the plaquette.
To describe these gauge transformations, we need to assign orientations to the links.
Here we have chosen the links to have positive directions along \(\SI{0}{\degree}\), \(\SI{120}{\degree}\), and \(\SI{240}{\degree}\) (measured from the \(x\)-axis), as indicated by black arrowheads in Fig.~\ref{fig:s3-triangular-lattice}.
Now if a link \(\lambda\) is oriented away from a vertex, then an \(S_3\) charge \(h\) located at this vertex will induce a gauge transformation via a left-multipliction, \(\ket{g}\mapsto \ket{hg}\).
Conversely, if \(\lambda\) points towards the vertex, the gauge transformation is a right multiplication \(\ket{g}\mapsto\ket{g h^{-1}}\).
In Kitaev \cite{KITAEV20032}, these operations define the left and right multiplication operators, \(M_+^h\ket{g}  = \ket{hg}\) and \(M_-^h\ket{g} = \ket{g h^{-1}}\), which obey the relations \(M_{\pm}^{h} M_{\pm}^{k} = M_{\pm}^{h k}\) and \([M_+^h, M_-^k] = 0\).\footnote{Note that we have changed the notation for these operators. Kitaev uses \(L_\pm^h\). We have opted for \(M_\pm^h\) to avoid confusion with the left-multiplied permutation matrices acting on the CGS coupling matrix, defined in the next section.}
In the spinor representation described in the previous section, \(M_+^h\) is represented by \(\ell(h)\) and \(M_-^h\) is represented by \(r(h)\), so that \({M_+^h}^{-1} \bsigma^z_\lambda {M_+^h} = \ell(h) \bsigma^z_\lambda\) and \({M_-^h}^{-1} \bsigma^z_\lambda {M_-^h} = r(h) \bsigma^z_\lambda \).
Combining the three spinors \(\bsigma_1^z, \bsigma_2^z, \bsigma_3^z\) around the plaquette into a big spinor \([\bsigma^z]\), 
\footnote{In general, the big spinors associated with the plaquette \(\pi\) will be indicated with a pair of brackets with the subscript \(\pi\) outside. Thus \([\bsigma^z]_\pi\) is the big gauge spinor on plaquette \(\pi\), and \([\mu^z]_\pi\) is the matter spinor \([\mu^z]_\pi\) on plaquette \(\pi\). In this section we are focusing only on one plaquette, so we drop the subscript for convenience.}
\begin{equation}
  [\bsigma^z] = 
  \begin{bmatrix}
    \bsigma^z_1 \\
    \bsigma^z_2 \\
    \bsigma^z_3
  \end{bmatrix} = 
  \begin{bmatrix}
    \sigma^z_{1, 1} \\
    \vdots \\
    \sigma^z_{3, 6}
  \end{bmatrix}\ ,
\end{equation}
we can represent the gauge transformations generated by the charges \(h, k, f\) located at the three vertices of the plaquette using the following block-diagonal matrices that multiply \(\bigspinor{\bsigma^z}\) on the left,
\begin{equation}
  \begin{split}
  R(h, 1, 1) & = \begin{bmatrix}
    r(h) & \\
    & \ell(h) \\
    & & I
  \end{bmatrix}\ , \\
  R(1, k, 1) & = \begin{bmatrix}
    I & \\
    & r(k) \\
    & & \ell(k)
  \end{bmatrix}\ , \\
  R(1, 1, f) & = \begin{bmatrix}
    \ell(f) \\
    & I \\
    & & r(f)
  \end{bmatrix}
\end{split}
\end{equation}
(see Fig.~\ref{fig:s3-plaquette-gauge-transformation}\subref{fig:s3-plaquette-gauge-transformation-single}).
The general gauge transformation can be written as a combination of the above matrices,
\begin{equation}
  \begin{split}
  R(h, k, f) & = R(h, 1,  1)\; R(1, k, 1)\; R(1, 1, f) \\
  & = \begin{bmatrix}[1.5]
    \ell(f)r(h) \\
    & \ell(h)r(k) \\
    & & \ell(k) r(f)
  \end{bmatrix}
\end{split}
\end{equation}
(see Fig.~~\ref{fig:s3-plaquette-gauge-transformation}\subref{fig:s3-plaquette-gauge-transformation-all}).
These gauge transformations form a group, which we call the \emph{local group of gauge transformations} (LGGT), denoted \(\mathcal{G}\).
The subgroup of \(\mathcal{G}\) consisting of elements of the form \(R(h, k, 1)\) are also of interest.
We call this subgroup the restricted LGGT, and denote it by \(\bar{\mathcal{G}}\).


\subsection{The \texorpdfstring{\(S_3\)}{S3} CGS interaction}\label{sec:s3-cgs-interaction}

Next, we introduce \(36\) matter spins, also grouped into a \(36\)-spinor,
\begin{equation}
  \bigspinor{\mu^z}= 
  \begin{bmatrix}
    \mu^z_{1} \\
    \mu^z_{2} \\
    \vdots \\
    \mu^z_{36}
  \end{bmatrix}\ .
\end{equation}
They are coupled to the matter spinors by
\begin{equation}
  H^\text{CGS} = - J_c \trans{[\mu^z]}\,W\,[\bsigma^z]
  \label{eq:s3-hamiltonian-coupling-term}
\end{equation}
via an interaction matrix
\begin{equation}
  W = 
  \begin{bmatrix}[1.5]
    \trans{v(1)} r(h_1) & \trans{v(1)} \ell(h_{1}) r(k_{1}) & \trans{v(1)} \ell(k_{1}) \\
    \trans{v(1)} r(h_2) & \trans{v(1)} \ell(h_{2}) r(k_{2}) & \trans{v(1)} \ell(k_{2}) \\
     & \vdots \\
    \trans{v(1)} r(h_{36}) & \trans{v(1)} \ell(h_{36}) r(k_{36}) & \trans{v(1)} \ell(k_{36})
  \end{bmatrix}\ .
  \label{eq:s3-w-matrix}
\end{equation}
\(W\) has \(36 \times 3\) entries of \(1\times 6\) vectors, if we see it as coupling \(36\) matter spins with \(3\) gauge \(6\)-spinors.
Alternatively, we can combine the three vectors in each row into a single \(18\)-entry row, and see \(W\) as a \(36\times 18\) matrix of \(\pm 1\) which couples \(36\) matter spins with the \(18\) physical spins that make up the gauge variables around a plaquette.
The \(a\)-th row of \(W\) is obtained by right-mutiplying a fixed initial row \(\begin{bmatrix}\trans{v(1)} & \trans{v(1)} & \trans{v(1)}\end{bmatrix}\) by the matrix \(R(h_a, k_a, 1) = \diag\left\{r(h_a), \ell(h_a)r(k_a), \ell(k_a)\right\}\) that represents the \(a\)-th element of the restricted LGGT \(\bar{\mathcal{G}}_p\).
Using the multiplication rules \eqref{eq:v-l-r-relations} and the fact that \(\trans{\ell(g)} = \ell(g^{-1})\), the product \(\trans{v(1)}\ell(h) r(k)\) can be rewritten as \(\trans{v(h^{-1}k)}\), so each entry of \(W\) is in fact a state label.
More precisely, each row of \(W\) is a distinct triple of state labels \(\begin{bmatrix} \trans{v(g_1)} & \trans{v(g_2)} & \trans{v(g_3)} \end{bmatrix}\) such that \(g_1 g_2 g_3 = 1\).
There are \(36\) such triples, since the constraint on their product implies that \(g_2 = g_1^{-1} g_3^{-1}\) while \(g_1\) and \(g_3\) can be taken as arbitrary elements of \(S_3\).
This provides another way of counting the number of rows of \(W\).

The key property of \(W\) is its invariance under the LGGT.
This means that given any element \(R(h, k, f)\) of the LGGT, there exists a permutation matrix \(L(h, k, f)\) such that
\begin{equation}
  \trans{L(h, k, f)}\; W\; R(h, k, f) = W \label{eq:w-invariance-full-lggt}\ .
\end{equation}
We prove this in Appendix~\ref{app:CGS-W-matrix-invariance} for a general gauge group \(G\) and illustrate the effect of one such pair of \(L(h, k, f)\) and \(R(h, k, f)\) in Appendix~\ref{app:s3-cgs-transformation-example}.
Once again, it is significant that \(L(h, k, f)\) is a permutation matrix, because physically it corresponds to a permutation of the matter spins \(\mu^z_{a}\).
The invariance condition \eqref{eq:w-invariance-full-lggt} means that, whenever the charges \(h, k,\) and \(f\) induce a gauge transformation on the gauge spinors, there is a corresponding permutation of the matter spins such that the joint effect of these two transformations on the gauge and matter spins will leave the coupling term \eqref{eq:s3-hamiltonian-coupling-term} invariant.

The combinatorial gauge symmetry of \(H^\text{CGS}\) preserves the gauge flux \(\phi(g_1, g_2, g_3) \equiv g_1 g_2 g_3\) through the plaquette.
Moreover, it can map any two configurations with the same flux to each other, that is, if \(\phi(g_1, g_2, g_3) = \phi(g_1', g_2', g_3')\), then there exists charges \(h, k,\) and \(f\) such that \((f g_1 h^{-1}, h g_2 k^{-1}, k g_3 f^{-1}) = (g_1', g_2', g_3')\).
We demonstrate this fact in Appendix~\ref{app:lggt-transitivity}.
This guarantees that any configuration of gauge variables that produces the same flux will have the same energy.

Even though the Hamiltonian \(H^\text{CGS}\) has the correct symmetry, it might in general still have accidental degeneracies that bring the energy of different flux sectors to the same level. 
In order to construct a system whose ground state satisfies the zero-flux condition, we need to shift the energies of different flux sectors away from each other, so that the zero-flux sector occupies the lowest energy levels.
This is achieved by applying a longitudinal field
\begin{equation}
  H^\text{flux} = J_f \sum_{a = 1}^{36} \mu^z_a\ ,
\end{equation}
where \(12 J_c < J_f < 18 J_c\).
This choice produces a gap of size up to \(8 J_c\) at \(J_f = 14 J_c\) between the \(\phi([\bsigma^z]) = 1\) state and the \(\phi([\bsigma^z])\neq 1\) states.
See Appendix~\ref{app:spectrum} for a detailed discussion of the spectrum and the choice of the field strength.

Finally, we have the Hamiltonian in equation \eqref{eq:cgs-hamiltonian-general}, which is invariant with respect to \(S_3\) gauge transformations, whose ground state consists only of states that satisfy the \(g_1 g_2 g_3 = 1\) constraint around each plaquette.

\section{General construction of non-Abelian combinatorial gauge theories}
\label{sec:general-construction}

Generalizing the above procedure, for any finite group \(G\), we can construct a two-body gauge invariant 2D lattice Hamiltonian with \(G\) as the gauge group.
On each link \(\lambda\) of the lattice, we place \(\abs{G}\) spins \(\sigma_{\lambda, j}^z\) to represent the gauge degrees of freedom.
Within a plaquette bounded by \(m\) links, we place \(N\) matter spins \(\mu_{\pi, a}^z\), where \(N = \abs{G}^{m - 1}\).
To simplify the notation, we combine the \(\abs{G}\) gauge spins on link \(\lambda\) into a spinor \(\bsigma^z_\lambda = \trans{[\sigma^z_{\lambda, 1}, \dots, \sigma^z_{\lambda, \abs{G}}]}\), denoted by a bold letter to distinguish it from its component spins.
We further group the \(m\) gauge spinors and \(N\) matter spins associated with plaquette \(\pi\) into spinors \([\bsigma^z]_\pi = \trans{[\bsigma_{\lambda_1}^z, \dots, \bsigma_{\lambda_m}^z]}_{\substack{\lambda_1, \dots \\\lambda_m \in \pi}}\) and \([\mu^z]_{\pi} = \trans{[\mu_{\pi, 1}^z, \mu_{\pi, 2}^z, \dots, \mu_{\pi, N}^z]}\), respectively.
The indices \(\lambda_i\) on the gauge spinors enumerate the links that surround the plaquette \(\pi\), and the double indices \((\pi, a)\) on the matter spins enumerate the matter spins within each plaquette.
The gauge-invariant Hamiltonian takes the same form as the \(S_3\) CGS Hamiltonian defined in \eqref{eq:H-as-sum-of-plaquettes},
\begin{equation}
  H = \sum_\pi \left( H^\text{proj}_\pi + H^\text{CGS}_\pi + H^\text{flux}_\pi\right)\ .\label{eq:cgs-hamiltonian-general}
\end{equation}
Just as in the \(S_3\) example, we will focus on one plaquette at a time and suppress the \(\pi\) index when we discuss each term of \eqref{eq:cgs-hamiltonian-general} in detail.

The first term of \eqref{eq:cgs-hamiltonian-general} projects the spins that make up the gauge spinor on each link into a \(\abs{G}\)-dimensional subspace that represents the gauge variables.
Since in a two dimensional lattice, each link belongs to two plaquettes, we have the relationship \(H^\text{proj}_\pi = \frac{1}{2}\sum_{\lambda\in\pi} H_\lambda^\text{proj}\), which allows us to rewrite the sum over plaquettes as a sum over links, \(\sum_\pi H^\text{proj}_\pi = \sum_\lambda H^\text{proj}_\lambda\).
On each link the projection term is an Ising interaction,
\begin{equation}
  H^\text{proj}_\lambda = J_p \left(\sum_{j = 1}^{\abs{G}} \sigma_{\lambda, j}^z - \abs{G} + 2\right)^2
  \label{eq:cgs-hamiltonian-projection-term}
\end{equation}
which is minimized when the components of the gauge spinor have magnetization \(\abs{G} - 2\).
(Since \(S_3\) is of order \(6\), the constant in the \(S_3\) projection term is \(\abs{G} - 2 = 4\), which brings it to the form in equation \eqref{eq:s3-hamiltonian-projection-term}.)
When \(J_p\) is sufficiently large, i.e. much greater than the gap of \(H^\text{CGS}\), the only relevent low energy states are those that satisfy \(\sum_{j} \sigma^z_{\lambda, j} = \abs{G} - 2\), which are spanned by states where exactly one of the spins \(\sigma^z_{\lambda, j}\) takes value \(-1\) while all others are \(+1\).
These states are then mapped to the states \(\ket{g_i}\) of the gauge variable, which correspond to the elements of the group \(G\).
This assumes an ordering of the elements of \(G\), and the \(i\)-th element \(g_i\) is mapped to the state where the \(i\)-th spin equals \(-1\).
We define \(\ket{g_j}\) by the rule \(\bsigma^z_{\lambda} \ket{g_j} = v(g_j)\ket{g_j}\), so the states are labeled by the value of the spinor \(\bsigma^z_\lambda\).
The state label \(v(g_j)\) is a vector whose entries are all \(+1\), except for the \(j\)-th entry, which is \(-1\). 
Using these basis vectors, we can construct representations matrices, \(\ell(h)\) and \(r(h)\), of left- and right-multiplication by a group element \(h\in G\), such that
\begin{align}
  \ell(h) v(g) & = v(h g) \ ,
  \label{eq:l-matrix-condition-general}\\
  r(h) v(g) & = v(g h^{-1}) \ .
  \label{eq:r-matrix-condition-general}
\end{align}
These matrices are the left- and right-regular representations of the group \(G\) \citep[p.~5]{fulton2013representation}.
Cayley's theorem states that when we left-multiply (or right-multiply) all elements of a group by a fixed element of the same group, the effect is the same as a permutation of the group elements.
Thus the matrices \(\ell(h)\) and \(r(h)\) are permutation matrices.
Since they also permute the eigenvalues of \(\sigma^z_{\lambda,j}\), they correspond to permutations of the underlying spins, which are physical symmetries of the system that preserve the commutators of the spin operators.
This implies that \(H^\text{proj}_\lambda\) is gauge invariant, since it is invariant under permutations of the component spins \(\sigma^z_{\lambda, j}\) of each spinor spinor \(\bsigma^z_\lambda\).
The matrices \(\ell(h)\) and \(r(h)\) bridge the abstract gauge theory and the physical realization by mapping the gauge symmetries acting on gauge variables that take value in abstract groups to physical symmetries of a system of spins.

A gauge charge of value \(h\) induces gauge transformations on the gauge variables situated on all neighboring links.
As in the \(S_3\) case, we need to orient all links in the lattice.
Given an orientation, a charge \(h\) located at a particular vertex induces a gauge transformation \(\bsigma^z_\lambda\mapsto \ell(h) \bsigma^z_\lambda\) if the link \(\lambda\) points away from the vertex, while if \(\lambda\) points towards the vertex, the gauge transformation is \(\bsigma^z_\lambda\mapsto r(h) \bsigma^z_\lambda\) instead.
At this point we zoom in on a fixed plaquette \(\pi\), whose boundary is formed by the links \(\lambda_1, \dots, \lambda_m\).
From here on we will suppress the index \(\pi\), with the understanding that the spinors \([\bsigma^z]\) and \([\mu^z]\), as well as the local group of gauge transformations \(\mathcal{G}\) and the coupling matrix \(W\), to be defined below, can all depend on \(\pi\) in the most general case.
Without loss of generality, suppose the links around this plaquette are all oriented counterclockwise, then a charge \(h\) at one of the vertices on the boundary induces \(r(h)\) on link entering the vertex and \(\ell(h)\) on the exiting link.
Thus we can write down a block diagonal matrix \(R_{i, i+1}(h)\) left-multiplying the spinor \([\bsigma^z]\) that represents the gauge transformation induced by a charge \(h\) situated between the \(i\)-th and \((i+1)\)-th link,
\begin{equation}
  \begin{split}
  & R_{i, i+1}(h)[\bsigma^z]\\
  & = \begin{bmatrix}[1.2]
    1 & \phantom{\sigma^z_\pi} \\ \phantom{r(h)} & \ddots & \phantom{\sigma^z_\pi} \\ & \phantom{\sigma^z_\pi} & r(h) & \phantom{\sigma^z_\pi} \\ & & \phantom{\sigma^z_\pi} & \ell(h) & \phantom{\sigma^z_\pi} \\ & & & \phantom{\sigma^z_\pi} & \ddots & \phantom{r(h)} \\ & & & & \phantom{\sigma^z_\pi} & 1
  \end{bmatrix}
  \begin{bmatrix}[1.2]
    {}\bsigma^z_{\lambda_1} \\
    \vdots \\
    \bsigma^z_{\lambda_i} \\
    \bsigma^z_{\lambda_{i+1}}\\
    \vdots \\
    \bsigma^z_{\lambda_m}
  \end{bmatrix}
  \\
  & =
  \begin{bmatrix}[1.3]
    {}\bsigma^z_{\lambda_1}\\
    \vdots \\
    r(h)\bsigma^z_{\lambda_i} \\
    \ell(h)\bsigma^z_{\lambda_{i+1}}\\
    \vdots \\
    \bsigma^z_{\lambda_m}
  \end{bmatrix}
\end{split}
  \label{eq:r-matrix-single-charge}
\end{equation}
A general gauge transformation can be expressed by a product of these matrices, 
\begin{equation}
R(h_1, \dots, h_m) = R_{1,2}(h_1)\; R_{2,3}(h_2)\;\dots \;R_{m,1}(h_m)\ .
\label{eq:r-matrix-general}
\end{equation}
Since \(\ell(h)\) commutes with \(r(h')\) for all \(h, h'\in G\), the matrices \(R_{i, i+1}(h_k)\) all commute with each other.
Moreover, the \(\ell\) and \(r\) matrices obey the group multiplication law, i.e., \(\ell(h) \ell(h') = \ell(h h')\) and \(r(h) r(h') = r(h h')\), so when we multiply \(R(h_1, \dots, h_m)\) with \(R(h_1', \dots, h_m')\), the result is \(R(h_1 h_1', \dots, h_m h_m')\).
Thus the gauge transformations represented by the matrices \(R(h_1, \dots, h_m)\) form a group, which we call the local group of gauge transformations (LGGT), denoted by \(\mathcal{G}\).
The subgroup where the last generator \(h_m = 1\) is called the restricted LGGT, denoted by \(\bar{\mathcal{G}}\).

The CGS coupling term \(H^\text{CGS}\) couples matter spins (combined into the \(N\) spinor \([\mu^z]\)) to the gauge spinors (combined into the \(m\abs{G}\)-spinor \([\bsigma^z]\)) on each plaquette,
\begin{equation}
  H^\text{CGS}_\pi = - J_c [\mu^z] W [\bsigma^z]
  \label{eq:general-cgs-term}
\end{equation}
The coupling matrix \(W\) is given by
\begin{widetext}
\begin{equation}
  W = 
  \begin{bmatrix}[1.5]
    \trans{v(1)} r(h_{1, 1}) & \trans{v(1)} \ell(h_{1, 1}) r(h_{1, 2}) & \trans{v(1)} \ell(h_{1, 2}) r(h_{1, 3}) & \cdots 
    & \trans{v(1)} \ell(h_{1, m-1}) \\
    \vdots & \vdots & \vdots & 
    & \vdots\\
    \trans{v(1)} r(h_{N, 1}) & \trans{v(1)} \ell(h_{N, 1}) r(h_{N, 2}) & \trans{v(1)} \ell(h_{N, 2}) r(h_{N, 3}) & \cdots 
    & \trans{v(1)} \ell(h_{N, m-1})
  \end{bmatrix}
\end{equation}
\end{widetext}
The first index \(a\) on \(h_{a, j}\) enumerates the elements of the restricted LGGT, so the number of rows of \(W\) and the number of matter spins are determined by the order of the restricted LGGT \(\tilde{\mathcal{G}}\).
Since each gauge transformation in the restricted LGGT is induced by \(m - 1\) charges, \(h_{a,1}, \dots, h_{a, m-1}\), the \((a, j)\)-th entry of \(W\) is right-multiplied by the charge \(h_{a, j}\) located between the \(j\)-th and \((j+1)\)-th link on the boundary of the plaquette, and left-multiplied by the charge \(h_{a, j-1}\) located beteen the \((j-1)\)-th and \(j\)-th link.
This matrix has combinatorial gauge symmetry, which means that given an arbitrary element of the LGGT, \(R(h_1, \dots, h_m)\), there exists a corresponding permutation matrix \(L(h_1, \dots, h_m)\), such that the following condition is satisfied,
\begin{equation}
  \trans{L(h_1, \dots, h_m)} \; W R(h_1, \dots, h_m) = W\ ,
  \label{eq:w-matrix-invariance-general-g}
\end{equation}
Physically, this means that the Hamiltonian \eqref{eq:general-cgs-term} possess a gauge symmetry which transforms the gauge variables by left- and right-multiplications induced by charges \(h_1, \dots, h_m\), and simultaneously permutes the matter spins according to ther permutation matrix \(L(h_1, \dots, h_m)\).
While we have only used the restricted LGGT to construct \(W\), it is invariant under the full LGGT.
We demonstrate this in Appendix.~\ref{app:CGS-W-matrix-invariance}.
The main consequence of this is that the number of matter spins necessary for the construction is \( |{\tilde{\mathcal{G}}}| = \abs{G}^{m -1}\), which is \(\abs{G}\) times smaller than a naive symmetrization with respect to all possible gauge transformations.


Just as in the \(S_3\) example and as demonstrated in Appendix.~\ref{app:lggt-transitivity}, the action of the LGGT is transitive among the different configurations of the gauge variable that produce the same flux through each plaquette.
This guarantees that each state has the same energy as its gauge symmetry partners, but leaves open the possibility that distinct flux sectors become degenerate.
Indeed, in the construction described above, the ground state of \(H^\text{proj} + H^\text{CGS}\) mixes flux sectors. 
In order to split this degeneracy and ensure that the ground state has zero flux, we need to apply a longitudinal field on the matter spins,
\begin{equation}
  H^\text{flux}_{\pi} = J_f \sum_{a = 1}^N \mu_{\pi, a}^z\ .
\end{equation}
This lowers the energy of the zero-flux states away from the non zero-flux states in the ground state manifold by up to \(8 J_c\) when \(J_f\) is between \((m\abs{G}-6)J_c\) and \(m\abs{G} J_c\).
The spectrum of \(H^\text{CGS}_\pi + H^\text{flux}_\pi\) is analyzed in detail in Appendix~\ref{app:spectrum}.

\section{Summary and outlook}


In this paper we have laid out a three-step process for constructing combinatorially gauge symmetric models that have exact non-Abelian gauge symmetries.
First, given a gauge group \(G\), we embed the gauge degrees of freedom in a physical system of spins, such that the left- and right-multiplication operators acting on gauge variables are represented by spin-flips and permutations of physical spins.
Second, we introduce an appropriate number of matter degrees of freedom to couple with the gauge degrees of freedom, and construct a coupling matrix from the orbit of an initial set of coupling constants under the action of the gauge symmetries, so that the resulting coupling matrix will satisfy an invariance condition of the form \eqref{eq:w-matrix-invariance-general-g}.
Third, if the ground state of the resulting Hamiltonian does not consist of the zero-flux states, we need to introduce additional gauge invariant terms or make modifications to the coupling term while preserving its gauge invariance, so that zero-flux states are shifted to the lowest energy.

Each step of the process is open to extensions beyond the ``standard'' method presented in Section~\ref{sec:general-construction}.
For the first step, alternative embeddings of gauge variables into physical degrees of freedom can occur for certain gauge groups, which may reduce the number of matter spins required, as in the case of the quaternion CGS theory discussed by \citet{Green2023}. 
In Appendix~\ref{app:gauge-embedding-general-requirements} we describe the mathematical constraints on such embeddings, which lead to a computational procedure for discovering them.
One such example is shown in Appendix~\ref{app:additional-examples}.
Following from generalized embeddings of gauge variables, alternative constructions of the CGS coupling and methods of flux-fixing are also possible.
We elaborate on these in Appendix~\ref{app:gauge-embedding-general-requirements}.

These two-body models that carry exact non-Abelian symmetries 
open up new avenues for exploring the intricate behavior of non-Abelian anyons and their associated phases using more realistic models.
Further investigation may shed light on the phases and the nature of excitations of these models, which should provide deeper insights into their relationship with standard predictions about non-Abelian anyons. 
Given the two-body nature of these models and their exact non-Abelian symmetries, they present a promising opportunity for realizing topologically ordered states on programmable quantum hardware. 

\section*{Acknowledgements}

The work of H.Y. and C.C. is supported by the DOE Grant DE-FG02-06ER46316.

\appendix



\section{Invariance of the coupling matrix in CGS models}
\label{app:CGS-W-matrix-invariance}

In non-Abelian gauge theories, fluxes are only defined up to conjugation, thus in general, a gauge transformation can change the value of the product \(g_1\cdots g_m\) of gauge variables around a plaquette, but cannot change the conjugacy class of this prouduct.
It is useful to distinguish the full local group of gauge transformation (LGGT) that preserves the flux up to conjugacy, \(\mathcal{G}\), from a restricted LGGT \(\bar{\mathcal{G}}\), which preserves the product \(g_1 \cdots g_m\) exactly.
The latter is a subgroup of the former.
This definition of the LGGT and the restricted LGGT is the same as the description given in Sec.~\ref{sec:general-construction}.
The gauge transformation represented by the \(R\) matrix \(R(h_1, \dots, h_{m-1}, 1)\) inserts the first \(m-1\) charges in the form \(h_i^{-1} h_i\) between the gauge variables in the product \(g_1\cdots g_m\), so the product is clearly left invariant.
On the other hand, the gauge transformation \(R(1, \dots, 1, h_m)\) conjugates the product \(g_1\cdots g_m\) by \(h_m\).
Thus the restricted LGGT as defined in Sec.~\ref{sec:general-construction} consist of precisely the gauge transformations that preserve the product \(g_1\cdots g_m\) exactly, and the full LGGT has additional elements that apply a conjugation to the product.
Any element \(R(h_1, \dots, h_{m-1}, h_m)\) of \(\mathcal{G}\) can be decomposed into an element \(R(h_1, \dots, h_{m-1}, 1)\) from \(\bar{\mathcal{G}}\) times an element of the form \(R(1, \dots, 1, h_m)\).
Therefore, when discussing the invariance of the coupling matrix \(W\), we can show it in two steps, first its invariance under \(\bar{\mathcal{G}}\), then its invariance under transformations of the form \(R(1, \dots, 1, h_m)\).


The rows of \(W\) are constructed out of an orbit under the action of the restricted LGGT by right multiplication of the matrices \(R(h_{a, 1}, \dots, h_{a, m - 1}, 1)\), so the \(a\)-th row of \(W\), denoted by \(W_{a\colinddummy}\), is
\begin{equation}
  \overbrace{\begin{bmatrix}\trans{v(1)} & \trans{v(1)} & \cdots & \trans{v(1)}\end{bmatrix}}^{m \text{ entries}} \; R(h_{a, 1}, \dots, h_{a, M-1}, 1)\ .
  \label{eq:w-row-definition}
\end{equation}
Hence, they also form an invariant subset under the same action.
This means that multiplying \(W\) on the right by \(R(h_1, \dots, h_{m-1}, 1) \in \bar{\mathcal{G}}\) preserves the set of rows as a whole, and only permutes the rows among themselves.
This means that the column operation given by the right-multiplication of \(R(h_1, \dots, h_{m-1}, 1)\) is equivalent to a row operation given by the left-multiplication of a permutation matrix, which we denote by \(L(h_1,\dots, h_{m-1}, 1)\),
\begin{equation}
  W\; R(h_1, \dots, h_{m - 1}, 1) = L(h_1, \dots, h_{m - 1}, 1) \; W\ .
  \label{eq:general-W-invariance-restricted-LGGT-two-sided}
\end{equation}
In other words, \(W\) is invariant under the joint action of \(L(h_1, \dots, h_{m - 1}, 1)\) and \(R(h_1, \dots, h_{m - 1}, 1)\),
\begin{equation}
  \trans{L(h_1, \dots, h_{m-1}, 1)}\; W\; R(h_1, \dots, h_{m-1}, 1) = W\ .\label{eq:general-W-invariance-restricted-LGGT}
\end{equation}
This demonstrate the invariance of \(W\) under the restricted LGGT.
Now we investigate the effect of gauge transformations of the form \(R(1, \dots, 1, h_m)\) on \(W\).
Right-multiplying the \(a\)-th row of \(W\) by \(R(1, \dots, 1, h_m)\) gives us 
\begin{widetext}
\begin{equation}
  W_{a\colinddummy}\; R(1, \dots, 1, h_m) = \\\begin{bmatrix}\trans{v(1)} r(h_{a, 1}) \ell(h_m) & v(1) \ell(h_{a, 1}) r(h_{a, 2})  \cdots & \trans{v(1)} \ell(h_{a, m - 1}) r(h_m) \end{bmatrix} \ .
  \label{eq:w-row-extra-gauge-transformation}
\end{equation}
\end{widetext}
Applying the conditions \eqref{eq:l-matrix-condition-general} and \eqref{eq:r-matrix-condition-general} to \(\ell(h_m)\), \(r(h_m)\), and \(v(1)\), we see that \(\ell(h_m) r(h_m) v(1) = v(h_m h_m^{-1}) = v(1)\).
Taking the conjugate of this identity, and making use of the fact that \(\trans{\ell(h)} = \ell(h)^{-1} = \ell(h^{-1})\) and \(\trans{r(h)} = r(h)^{-1} = r(h^{-1})\), we get \(\trans{v(1)} = \trans{v(1)} \ell(h_m^{-1}) r(h_m^{-1})\).
This identity can be used to insert \(\ell(h_m^{-1}) r(h_m^{-1})\) after \(\trans{v(1)}\) in every entry of \eqref{eq:w-row-extra-gauge-transformation}.
The first and last entries become \(\trans{v(1)} \ell(h_m^{-1}) r(h_m^{-1}) r(h_{a, 1}) \ell(h_m) = \trans{v(1)} r(h_m^{-1} h_{a, 1})\) and \(\trans{v(1)} \ell(h_m^{-1}) r(h_m^{-1}) \ell(h_{a, m - 1}) r(h_m) = \ell(h_m^{-1} h_{a, m - 1})\), while the other entries become \(\trans{v(1)} \ell(h_m^{-1}) r(h_m^{-1}) \ell(h_{a, i + 1}) r(h_{a, i}) = \trans{v(1)} \ell(h_m^{-1} h_{a, i + 1}) r(h_m^{-1} h_{a, i})\).
Combining these results, we get
\begin{widetext}
\begin{equation}
\begin{aligned}
  W_{a\colinddummy}\; R(1, \dots, 1, h_m) & = \begin{bmatrix}
    \trans{v(1)} r(h_m^{-1} h_{a, 1}) & \trans{v(1)} \ell(h_m^{-1} h_{a, 1}) r(h_m^{-1} h_{a, 2}) & \cdots & \trans{v(1)} \ell(h_m^{-1} h_{a, m - 1})
  \end{bmatrix} \\
  & = \begin{bmatrix}
    \trans{v(1)} & \trans{v(1)} \cdots & \trans{v(1)}
  \end{bmatrix}
  R(h_m^{-1} h_{a, 1}, \dots, h_m^{-1} h_{a, m - 1}, 1) \\
  & = \begin{bmatrix}
    \trans{v(1)} & \trans{v(1)} \cdots & \trans{v(1)}
  \end{bmatrix} R(h_m^{-1} h_{a, 1} h_m, \dots, h_m^{-1} h_{a, m-1} h_m, 1)\; R(h_m^{-1}, \dots, h_m^{-1}, 1)\ .
\end{aligned}
\label{eq:w-row-extra-gauge-transformation-intermediate}
\end{equation}
\end{widetext}
Taking the first two factors in the product together, we get
\[
W_{a\colinddummy}\; R(h_m^{-1}h_{a,1}h_m, \dots, h_m^{-1}h_{a, m-1} h_{m, 1})\ ,
\]
which is analogous to the definition of \(W_{a\colinddummy}\) in equation \eqref{eq:w-row-definition}, except that instead of \(R(h_{a, 1}, \dots, h_{a, m - 1}, 1)\), we have conjugated each of these elements of \(\bar{\mathcal{G}}\) by \(R(h_m, \dots, h_m, 1)\). 
Since conjugating all group elements by a fixed element is an automorphism, that is, it permutes the elements of \(\bar{\mathcal{G}}\) but leaves the whole set invariant, this alteration is equivalent to a permutation of the rows of \(W\), which again corresponds to a left-multiplication by a permutation matrix which we will denote by \(C(h_m)\).
In other words,
\begin{multline}
  \begin{bmatrix}
    \trans{v(1)} & \cdots & \trans{v(1)}
  \end{bmatrix}
  R(h_m^{-1} h_{a, 1} h_m, \dots, h_m^{-1} h_{a, m - 1} h_m, 1)
  \\
  \begin{aligned}
    & = \sum_{b} C(h_m)_{ab} \begin{bmatrix}
      \trans{v(1)} & \trans{v(1)} & \cdots & \trans{v(1)}
    \end{bmatrix} \\
    & \qquad
    \times R(h_{b, 1} h_m, \dots, h_{b, m - 1}, 1) \\
    & = \sum_{b} C(h_m)_{ab}\; W_{b\colinddummy}
  \end{aligned}
\end{multline}
Applying \(R(h_m^{-1}, \dots, h_m^{-1}, 1)\) to both sides of this identity to get back to the right hand side of \eqref{eq:w-row-extra-gauge-transformation-intermediate}.
Making use of \eqref{eq:general-W-invariance-restricted-LGGT-two-sided}, we then get
\begin{equation}
\begin{split}
  & \quad W_{a\colinddummy}\; R(1, \dots, 1, h_m) \\
  & = \sum_{b} C(h_m)_{ab}\; W_{b\colinddummy}\; R(h_m^{-1}, \dots, h_m^{-1}, 1) \\
  & = \sum_{b, c}\; C(h_m)_{ab}\; L(h_m^{-1},\dots, h_m^{-1}, 1)_{bc}\; W_{c\colinddummy}\ .
\end{split}
\end{equation}
When we assemble these relations between the rows of \(W\) back into the full matrix, this identity becomes
\begin{equation}
  W\; R(1, \dots, 1, h_m) = C(h_m)\; L(h_m^{-1}, \dots, h_m^{-1}, 1)\; W
\end{equation}
Thus if we define \(L(1, \dots, 1, h_m)\equiv C(h_m)\; L(h_m^{-1}, \dots, h_m^{-1}, 1)\), we get the invariance condition 
\begin{equation}
  \trans{L(1, \dots, 1, h_m)}\; W\; R(1, \dots, 1, h_m) = W\ .
\end{equation}
This allows us to extend the invariance condition \eqref{eq:general-W-invariance-restricted-LGGT} to
\begin{equation}
  \trans{L(h_1, \dots, h_m)}\; W\; R(h_1, \dots, h_m) = W\ ,
\end{equation}
by defining \(L(h_1, \dots, h_{m-1}, h_m) \equiv L(h_1, \dots, h_{m-1}, 1)\; L(1, \dots, 1, h_m)\), corresponding to the factorization \(R(h_1, \dots, h_{m-1}, h_m) = R(h_1, \dots, h_{m-1}, 1)\; R(1, \dots, 1, h_m)\).
This means that \(W\) is in fact invariant under the full LGGT.

\section{Example combinatorial gauge symmetry transformation of the \texorpdfstring{\(S_3\)}{S3} CGS coupling term}
\label{app:s3-cgs-transformation-example}

\begin{figure*}[tb]
  \centering
  \includegraphics[width=0.8\textwidth]{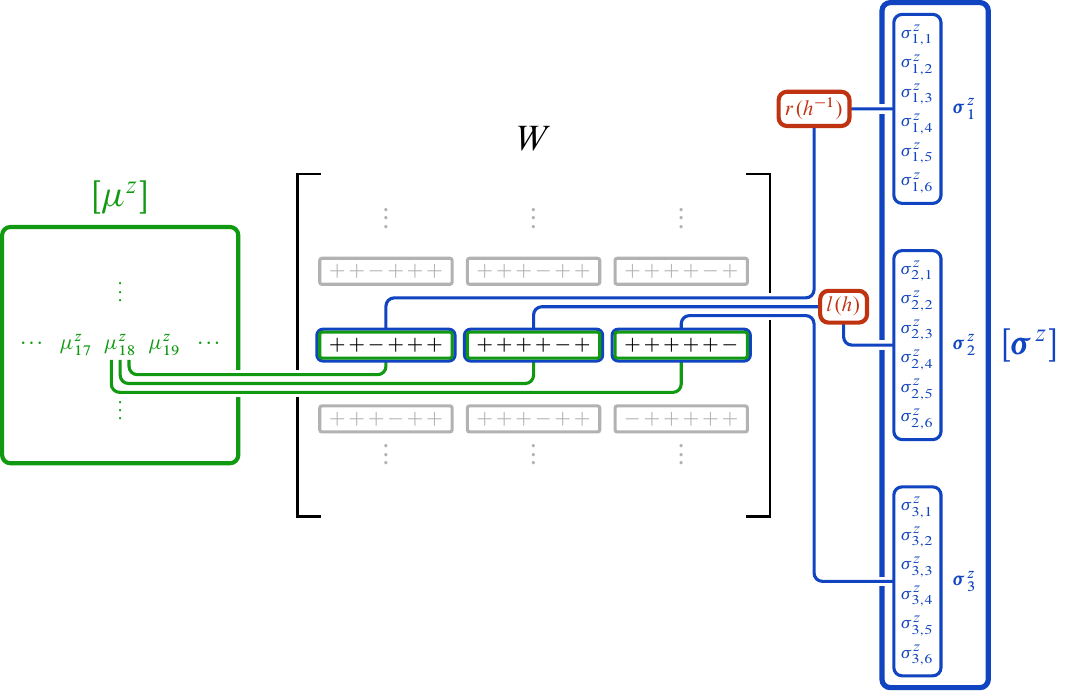}
  \caption{An illustration of the structure of the \(S_3\) CGS interaction term. The Hamiltonian \eqref{eq:s3-hamiltonian-coupling-term} can be understood at three levels of abstraction. At the lowest level of physical spins, it couples individual spins \(\sigma^z_{\lambda, i}\) with \(\mu^z_a\) through and Ising interaction with strength \(J_c\). Abstracting from this, the spins \(\sigma^z_{\lambda, i}\) are grouped into spinors \(\boldsymbol{\sigma}^z_\lambda\) representing the gauge variables, which coupled to \(\mu^z_a\) via \(1\times 6\)-vector coupling constants. These terms are illustrated by the blue and green lines connecting gauge spinors through the entries of \(W\). At the highest level of abstraction, it can be regarded as a coupling between the spinors \([\boldsymbol{\sigma}^z]\) and \([\mu^z]\). When a gauge transformation is applied to the gauge spinors, such as the one in Fig.~\ref{fig:s3-plaquette-gauge-transformation-single}, left- and right-multiplication matrices \(r(h^{-1})\) and \(l(h)\), represented in red, are inserted between the \(1\times 6\)-vector entries of \(W\) and the gauge spinors \(\boldsymbol{\sigma}^z_1\). These matrices can be applied to the right on the gauge spinors, or to the left on the coupling matrix \(W\). Transfering from the former to the latter point of view is the first step in our argument establishing the combinatorial gauge symmetry.}
  \label{fig:s3-w-matrix-structure}
\end{figure*}

Here we illustrate the invariance condition \eqref{eq:w-invariance-full-lggt} for the \(S_3\) CGS coupling matrix \(W\) described in Sec.~\ref{sec:s3-cgs-interaction} with an explicit example.

First, we present the explicit form of \(W\)
\setcounter{MaxMatrixCols}{18}
\begin{equation}
  W = 
  \begingroup
  \setlength\arraycolsep{1pt}
  \def\arraystretch{0.8}
  \begin{bmatrix+}{6|6|6}
    - & + & + & + & + & + & - & + & + & + & + & + & - & + & + & + & + & +\\
    \hline
    - & + & + & + & + & + & + & - & + & + & + & + & + & + & - & + & + & +\\
    \hline
    - & + & + & + & + & + & + & + & - & + & + & + & + & - & + & + & + & +\\
    \hline
    - & + & + & + & + & + & + & + & + & - & + & + & + & + & + & - & + & +\\
    \hline
    - & + & + & + & + & + & + & + & + & + & - & + & + & + & + & + & - & +\\
    \hline
    - & + & + & + & + & + & + & + & + & + & + & - & + & + & + & + & + & -\\
    \hline
    + & - & + & + & + & + & + & + & - & + & + & + & - & + & + & + & + & +\\
    \hline
    + & - & + & + & + & + & - & + & + & + & + & + & + & + & - & + & + & +\\
    \hline
    + & - & + & + & + & + & + & - & + & + & + & + & + & - & + & + & + & +\\
    \hline
    + & - & + & + & + & + & + & + & + & + & - & + & + & + & + & - & + & +\\
    \hline
    + & - & + & + & + & + & + & + & + & + & + & - & + & + & + & + & - & +\\
    \hline
    + & - & + & + & + & + & + & + & + & - & + & + & + & + & + & + & + & -\\
    \hline
    + & + & - & + & + & + & + & - & + & + & + & + & - & + & + & + & + & +\\
    \hline
    + & + & - & + & + & + & + & + & - & + & + & + & + & + & - & + & + & +\\
    \hline
    + & + & - & + & + & + & - & + & + & + & + & + & + & - & + & + & + & +\\
    \hline
    + & + & - & + & + & + & + & + & + & + & + & - & + & + & + & - & + & +\\
    \hline
    + & + & - & + & + & + & + & + & + & - & + & + & + & + & + & + & - & +\\
    \hline
    + & + & - & + & + & + & + & + & + & + & - & + & + & + & + & + & + & -\\
    \hline
    + & + & + & - & + & + & + & + & + & - & + & + & - & + & + & + & + & +\\
    \hline
    + & + & + & - & + & + & + & + & + & + & - & + & + & + & - & + & + & +\\
    \hline
    + & + & + & - & + & + & + & + & + & + & + & - & + & - & + & + & + & +\\
    \hline
    + & + & + & - & + & + & - & + & + & + & + & + & + & + & + & - & + & +\\
    \hline
    + & + & + & - & + & + & + & - & + & + & + & + & + & + & + & + & - & +\\
    \hline
    + & + & + & - & + & + & + & + & - & + & + & + & + & + & + & + & + & -\\
    \hline
    + & + & + & + & - & + & + & + & + & + & - & + & - & + & + & + & + & +\\
    \hline
    + & + & + & + & - & + & + & + & + & + & + & - & + & + & - & + & + & +\\
    \hline
    + & + & + & + & - & + & + & + & + & - & + & + & + & - & + & + & + & +\\
    \hline
    + & + & + & + & - & + & + & + & - & + & + & + & + & + & + & - & + & +\\
    \hline
    + & + & + & + & - & + & - & + & + & + & + & + & + & + & + & + & - & +\\
    \hline
    + & + & + & + & - & + & + & - & + & + & + & + & + & + & + & + & + & -\\
    \hline
    + & + & + & + & + & - & + & + & + & + & + & - & - & + & + & + & + & +\\
    \hline
    + & + & + & + & + & - & + & + & + & - & + & + & + & + & - & + & + & +\\
    \hline
    + & + & + & + & + & - & + & + & + & + & - & + & + & - & + & + & + & +\\
    \hline
    + & + & + & + & + & - & + & - & + & + & + & + & + & + & + & - & + & +\\
    \hline
    + & + & + & + & + & - & + & + & - & + & + & + & + & + & + & + & - & +\\
    \hline
    + & + & + & + & + & - & - & + & + & + & + & + & + & + & + & + & + & -
  \end{bmatrix+}
  \endgroup
\end{equation}
Each \(1\times 6\) submatrix, separated by lines in the representation above, corresponds to an entry of \eqref{eq:s3-w-matrix} that couples a single matter spin to a single gauge spinor.
In each row we have three such entries, which makes up \(18\) columns.
The \(36\) rows of \(W\) are determined by the \(36\) elements of the restricted LGGT \(\bar{\mathcal{G}}\).
To understand the structure of this matrix in terms of the gauge spinors, see Fig.~\ref{fig:s3-w-matrix-structure}.

\begin{figure*}[tb]
  \centering
  \includegraphics[width=0.8\textwidth]{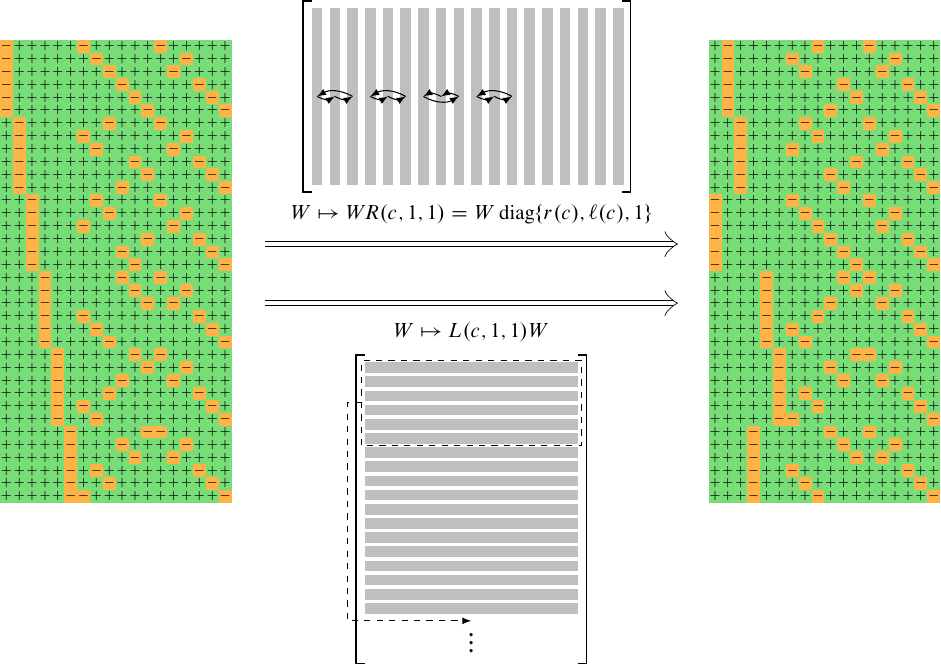}
  \caption{An example of a symmetry of the matrix \(W\) in the gauge-matter coupling term of an \(S_3\) combinatorial gauge theory. The positive and negative entries of \(W\) are colored in green and yellow to aid visualization. The column transformation represented by the diagram above the top arrow achieves the same effect as the row transformation represented by the diagram on the bottom.}
  \label{fig:s3-w-matrix-transformation-example}
\end{figure*}

Next we right-multiply \(W\) by the block-diagonal matrix 
\begin{equation}
  R(c, 1, 1) = \begin{bmatrix}
    r(c) \\
    & \ell(c) \\
    & & 1
  \end{bmatrix}\ ,
\end{equation}
which corresponds to the gauge transformation induced by a gauge charge of value \(c\) located on the vertex between link \(1\) and \(2\) of the triangular plaquette (see Fig.~\ref{fig:s3-plaquette-gauge-transformation}\subref{fig:s3-plaquette-gauge-transformation-single} and take \(h = c\)).
From equation \eqref{eq:s3-r-operator-generators} we see that when \(r(c)\) is multiplied on the right of the first six columns of \(W\), it permutes the first three columns cyclically according to \(\left(\begin{smallmatrix} 1 & 2 & 3 \\ 2 & 3 & 1\end{smallmatrix}\right)\) and the next three columns according to \(\left(\begin{smallmatrix} 4 & 5 & 6 \\ 5 & 6 & 4 \end{smallmatrix}\right)\).
Similarly, \(\ell(c)\) permutes the columns \(7\) through \(9\) as \(\left(\begin{smallmatrix}7 & 8 & 9 \\ 9 & 7 & 8\end{smallmatrix}\right)\) and the columns \(10\) through \(12\) as \(\left(\begin{smallmatrix}10 & 11 & 12 \\ 11 & 12 & 10\end{smallmatrix}\right)\).
The remaining columns of \(W\) are left changed by \(R(c, 1, 1)\).
This is illustrated in Fig.~\ref{fig:s3-w-matrix-transformation-example} by the diagram on the top.

Inspecting \(W\; R(c, 1, 1)\) closely, we see that its rows are identical to those in \(W\), but shuffled in groups of six.
Specifically, if we take the first six rows of \(W\) and move them below row \(18\), and move the rows \(19\) through \(24\) to bottom of the matrix, we obtain \(W\; R(c, 1, 1)\).
This is represented by the diagram on the bottom of Fig.~\ref{fig:s3-w-matrix-transformation-example}
This means that if we define \(L(c, 1, 1)\) as the permutation matrix that corresponds to these permutations, we can establish the relation \(W\; R(c, 1, 1) = L(c, 1, 1)\; W\).
Since permutation matrices are orthogonal, multiplying both sides by \(\trans{L(c, 1, 1)}\) gives us an instance of the invariance relaton \eqref{eq:w-invariance-full-lggt} where \(h = c\) and \(k = f = 1\),
\begin{equation}
  \trans{L(c, 1, 1)}\; W\; R(c, 1, 1) = W\ .
\end{equation}

\section{Transitivity of the action of the LGGT}
\label{app:lggt-transitivity}
So far we have described gauge transformations in terms of preserving products of the form \(g_1 g_2 \cdots g_m\).
The most general form of such transformation maps \(g_i\) to \(h_{i-1}\, g_i\, h_i\) for \(2\leq i \leq m - 1\) with arbitrary \(h_i\)'s, while \(g_1\) and \(g_m\) are mapped to \(g_1 h_1^{-1}\) and \(h_{m - 1}\, g_m\), respectively.
A combinatorial gauge symmetric model utilizes these transformations to construct an interaction that has the same gauge symmetries as a flux term \(g_1 g_2 \cdots g_m\).
One requirement for the resulting system is that, whenever two configurations of the gauge variables, \(g_1, \dots, g_m\) and \(g_1', \dots, g_m'\), have the same flux, i.e. \(g_1\cdots g_m = g_1' \cdots g_m'\), there should be states that have the same energies and these exact configurations of the gauge variables.
This can be guaranteed if the restricted LGGT acts \emph{transitively} on the set of configurations of gauge variables that have the same flux.
In other words, we would like to show that if \(g_1 g_2 \cdots g_m = g_1' g_2' \cdots g_m'\), there exists a set of charges \(h_1, \dots, h_{m-1}\) such that \(g_1' = g_1 h_1^{-1}, g_2' = h_1 g_2 h_2^{-1}, \dots, g_m' = h_{m-1}\, g_m\).
The charges \(h_1, \dots, h_{m-1}\) can in fact be solved from the first \(m-1\) identities here, because the first only contains \(h_1\), and each following one relates \(h_i\) to \(h_{i - 1}\).
To be precise, we have \(h_1 = g_1^{\prime-1} g_1\), and \(h_i = g_i^{\prime-1} h_{i-1}\; g_i\) for \(2 \leq i \leq m - 1\).
Hence, we need only prove that the last condition \(g_m' = h_{m-1}\, g_m\) is consistent with this set of charges.
Plugging the expression for \(h_{i-1}\) into that of \(h_i\) successively, we get the following result for \(h_{m-1}\),
\begin{equation}
\begin{split}
  h_{m-1} & = g_{m-1}^{\prime-1}\, g_{m-2}^{\prime-1} \cdots g_1^{\prime-1}\, g_1\, g_2 \cdots g_{m-1} \\
  & = (g_1'\, g_2' \cdots g'_{m-1})^{-1}\, g_1\, g_2 \cdots g_{m-1}
\end{split}
\end{equation}
Rearranging \(g_1'\, g_2'\dots g_{m-1}'\, g_m' = g_1\, g_2 \dots g_{m-1}\, g_m\), we get 
\begin{equation}
  (g_1'\, g_2' \cdots g'_{m-1})^{-1}\, g_1\, g_2 \cdots g_{m-1} = g_m'\, g_m^{-1}\ .
\end{equation}
Thus \(h_{m-1}\, g_m = g_m'\, g_m^{-1}\, g_m = g_m'\).

\section{Spectrum of the \texorpdfstring{\(S_3\)}{S3} plaquette Hamiltonian}\label{app:spectrum}

Here we first study the spectrum of \(H^\text{CGS}\) in the \(H^\text{proj} = 0\) subspace, focusing on the lowest energy state in each flux sector.
Gauge invariance ensures all states related by gauge symmetries have the same energy.
These states have the same flux and their matter spin configurations are related by a permutation.
Thus we need only consider one representative of every flux sector and minimize the energy with respect to the matter spins.
Here we introduce the notation \(\varphi([\boldsymbol{\sigma}^z])\) for the total flux threaded through the plaquette when the gauge spinors take the value \([\boldsymbol{\sigma}^z]\).
Thus for the \(\varphi([\boldsymbol{\sigma}^z]) = g\) states, we can choose the representative \([\bsigma^z]\) to \(\trans{[v(1), v(1), v(g)]}\) and minimize \(E_J = -J_c\trans{[\mu^z]} W[\bsigma^z] = - J_c \sum_{a}\mu^z_a \sum_i W_{ai}\bsigma^z_i\) with respect to \(\mu^z_a\)'s, which will give us the ground state energy of the flux-\(g\) sector.
Using the fact that \(\trans{v(g)} v(h) = 4\delta_{gh} + 2\), we can enumerate all the possible values of \(\sum_i W_{ai} \sigma^z_i\) and count their multiplicities.
In the \(\varphi([\boldsymbol{\sigma}^z]) = 1\) sector, we get
\begin{equation}
  E_1 = - 18 J_c \sum_{a = 1}^{1} \mu^z_{\perm_a} - 10 J_c \sum_{a = 2}^{16} \mu^z_{\perm_a} - 6 J_c \sum_{a = 17}^{36} \mu^z_{\perm_a}\ ,
\end{equation}
where \(\perm\) is a permutation of the \(36\) indices.
In all other sectors where \(\varphi([\boldsymbol{\sigma}^z]) \neq 1\), the effective Hamiltonian is
\begin{equation}
  E_g = - 14 J_c \sum_{a = 1}^{3} \mu^z_{\perm_a} - 10 J_c \sum_{a = 4}^{15} \mu^z_{\perm_a} - 6 J_c \sum_{a = 16}^{36} \mu^z_{\perm_a}\ .
\end{equation}
When \(\mu^z_a = +1\) for all \(a\)'s, both \(E_1\) and \(E_g\) achieve a minimum of \(- 288 J_c\).
However, when a uniform longitudinal field of strength \(14J_c\) is applied, the effective Hamiltonians become
\begin{equation}
  E_1 + H^\text{flux} = - 4 J_c \sum_{a = 1}^{1} \mu^z_{\perm_a} + 4 J_c \sum_{a = 2}^{16} \mu^z_{\perm_a} + 8 J_c \sum_{a = 17}^{36} \mu^z_{\perm_a}\ .
\end{equation}
and
\begin{equation}
  E_g + H^\text{flux} = 0 J_c \sum_{a = 1}^{3} \mu^z_{\perm_a} + 4 J_c \sum_{a = 4}^{15} \mu^z_{\perm_a} + 8 J_c \sum_{a = 16}^{36} \mu^z_{\perm_a}\ .
\end{equation}
The former is minimized at \(- 224 J_c\) when one of \(\mu^z_a\) is positive and all others are negative, and the latter is minimized at \(- 216 J_c\) when all but three of \(\mu^z_a\) are negative, while the remaining three can be either positive or negative.
The maximum gap in the \(S_3\) case is \(8J_c\), which we will see is indepedent of \(G\) and the lattice configuration number. 

The above arguments can be extended to the general non-Abelian case.
Here we refer to the standard construction described in Sec.~\ref{sec:general-construction}, where \(v(g)_i = 1 - 2\delta_{g, g_i}\).
In this case \(\trans{v(g)} v(h) = \delta_{gh}\abs{G} + (1 - \delta_{gh}) (\abs{G} - 4)\), which is always positive for a non-abelian group, where \(\abs{G} > 4\).
Thus for a fixed \([\sigma^z]\), in the effective Hamiltonian
\begin{equation}
\begin{split}
  E_J & = - J_c \trans{[\mu^z]} W [\sigma^z] =  - J_c \sum_{a = 1}^{N} \eta_a([\sigma^z]) \mu^z_a \\
  & =  - J_c \sum_{a = 1}^{N} \left(\sum_{i = 1}^{k} W_{ai} \sigma^z_i\right) \mu^z_a 
\end{split}
\end{equation}
the coefficient \(\eta_a([\sigma^z])\) is always positive.
When \(\varphi([\boldsymbol{\sigma}^z]) = 1\), using the representative configuration \([\boldsymbol{\sigma}^z] = \trans{[v(1), \dots, v(1)]}\), we see that the maximum of \(\eta_a([\sigma^z])\) is \(k\abs{G}\), achieved when \(W_{ai} = v(1)\) for all \(i = 1, \dots, k\).
When \(\phi([\sigma^z]) = g \neq 1\), we can take the representative configuration \([\sigma^z] = \trans{[v(g), v(1), \dots, v(1)]}\) and get a maximum of \(\eta_a([\sigma^z]) = k\abs{G} - 4\), once again when \(W_{ai} = v(1)\) for all \(i\).
On the other hand, the minimum of \(\eta_a([\sigma^z])\) is \(k(\abs{G} - 4)\) with any flux \(\phi([\sigma^z])\), achieved when \(W_{ai}\neq \sigma^z_i\) for all \(i\).
Next we apply an additional longtitudinal field \(H^\text{flux} = J_c(k\abs{G} - 4) \sum_a \mu^z_a\) is applied to the matter spins, so the effective Hamiltonian becomes
\begin{equation}
\begin{split}
E_J + H^\text{flux} & = - J_c \sum_{a = 1}^{N} \eta'_a([\sigma^z]) \mu^z_a \\
& = - J_c \sum_{a = 1}^{N} (\eta_a([\sigma^z_a]) - (k\abs{G} - 4)) \mu^z_a
\end{split}
\end{equation}
where \(\eta'_a([\sigma^z])\) is negative, except for a single term where \(W_a = \trans{[\sigma^z]}\), in which case \(\eta'_a = 4\).
Now to minimize \(E_J + H^\text{flux}\), we choose the sign of \(\mu^z_a\) according to the sign of \(\eta'_a\).
The preceeding argument about the sign of \(\eta'_a([\sigma^z])\) shows that when \(\phi([\sigma^z]) = g \neq 1\), the minimum energy is
\begin{equation}
\begin{split}
  E_g & = - J_c \sum_{a = 1}^{N} (- \eta'_a([\sigma^z])) \\
  & = J_c \sum_{a = 1}^{N} \eta_a([\sigma^z]) - J_c N (k\abs{G} - 4)\ ,
\end{split}
\end{equation}
achieved when all \(\mu^z_a\) are negative.
When \(\phi([\sigma^z]) = 1\), the minimum energy is
\begin{equation}
\begin{split}
  E_1 & = - J_c \sum_{a = 1}^{N} (- \eta'_a([\sigma^z])) - 8 J_c \\
  & = J_c \sum_{a = 1}^{N} \eta_a([\sigma^z]) - J_c N (k\abs{G} - 4) - 8 J_c\ ,
\end{split}
\end{equation}
achieved when all but one of \(\mu^z_a\) are negative.
To compare \(E_g\) and \(E_1\), we need to know the value of \(\sum_a \eta_a([\sigma^z]) = \sum_a \sum_i W_{ai} \sigma^z_i\) in the various flux sectors.
This sum is in fact independent of \([\sigma^z]\).
To see this, first swap the order of the sums and consider the sum over \(a\) for a fixed \(\sigma^z_i = v(g)\).
We can rewrite \(\sum_a W_{ai} \sigma^z_i\) as
\begin{equation}
\begin{split}
  \sum_a \trans{v(h_a)} v(g) & = \sum_a \left(4\delta_{h_a g} + (\abs{G} - 4)\right) \\
  & = \sum_a \left(4\delta_{h_a g^{-1},1} + (\abs{G} - 4)\right)\ .
\end{split}
\end{equation}
Recall that \(W\) is constructed from an orbit of the restricted LGGT, which is isomorphic to a direct product of \(k - 1\) copies of \(G\), so in any column of \(W\), each element of \(G\) appears exactly \(\abs{G}^{k - 2}\) times.
Since right-multiplying by \(g^{-1}\) only permutes the elements of \(G\), the expression \(h_a g^{-1}\) ranges over the same elements regardless of the value of \(g\).
Therefore, the value of \(\sum_a \eta([\sigma^z])\) is independent of \([\sigma^z]\), which means that \(E_g\) is always \(8J_c\) greater than \(E_1\), corroborating the result we obtained from explicit computation in the \(S_3\) case.

\section{General requirements for representations of gauge variables}
\label{app:gauge-embedding-general-requirements}

The Hilbert space \(\mathcal{H}_G\) for gauge degree of freedom is spanned by states that correspond to elements of the gauge group \(\{\ket{g}: g\in G\}\).
Such a variable is located on one oriented edge of the lattice and undergoes gauge transformations induced by charges located at either end points of the edge.
A charge \(h\) at the starting vertex induces a gauge transformation \(\ket{g}\mapsto \ket{hg}\), and \(h\) located at the ending vertex induces \(\ket{g}\mapsto \ket{gh^{-1}}\).
These two types of symmetry operations define the left- and right-multiplication operators \(M^h_+\) and \(M^h_-\), respectively.
The set of left-multiplication operators, denoted by \(M_+(G)\), is isomorphic to the group \(G\) itself, and the same is true for the set of right-multiplication operators \(M_-(G)\).
Together, these operators generate \(M(G)\), the group of multiplication operators.
A physical realization of a gauge degree of freedom needs to embed the Hilbert space \(\mathcal{H}_G\) in the Hilbert space \(\mathcal{H}\) of some physical degrees of freedom.
Moreover, these physical degrees of freedom need to possess a symmetry group that contains the gauge transformations \(M(G)\) as a subgroup.

Before we set out to find such embeddings, we need to clarify the structure of \(M(G)\).
We have already noted that \(M(G)\) contains two isomorphic copies of \(G\).
However, while left- and right-multiplication operators commute, i.e. \([M_+^g, M_-^h] = 0\) for all \(h, g\in G\), this does not imply that the group generated by \(M_+(G) \cup M_-(G)\) is isomorphic to \(G\times G\), because the intersection of the two sets is non-empty.
The elements that lie in the intersection correspond to the elements in the center of the gauge group, \(Z(G) = \{h\in G: h g = g h, \forall g \in G\}\).
To see this, note that \(hg = gh\) for all \(g\) implies that \(M^h_+ = M_-^{h^{-1}}\), or \(M^h_+ M^h_- = I\).
Thus if we identify \(M^h_+\) with \((h, 0)\in G\times G\) and \(M^{h}_-\) with \((0, h)\), the condition \(M^h_+ M^h_- = I\) means that the group generated by \(M_+(G) \cup M_-(G)\) is really \(G\times G / \{(h, h): h \in Z(G)\}\).
Defining the diagonal map \(\Delta: G\to G\times G\) as \(g\mapsto (g, g)\), we can write this as \(M(G)\iso G\times G / \Delta(Z(G))\).

Thus the general procedure for searching for an embedding of gauge variables is as follows.
Given a gauge group \(G\), we first compute the group of multiplication operators \(M(G)\).
Choosing a physical system with a symmetry group \(S\) acting on its Hilbert space \(\mathcal{H}\), we search among subgroups of \(S\) for an isomorphic copy of \(M(G)\).
For small groups \(G\), this can be done with computer algebra systems such as GAP \cite{GAP4}.
If an embedding of \(M(G)\) into \(S\) does not exist, we increase the system size or change the physical degrees of freedom and search again.
When an embedding is found, we search among normal subgroups of \(M(G)\) that are isomorphic to \(G\) and identify two such subgroups that intersect precisely at their centers.
These groups will be identified with \(M_+(G)\) and \(M_-(G)\), respectively.
Next we need to choose a state in \(\mathcal{H}\) that corresponds to the identity element of the gauge group, which must satisfy \(M_+^h M_-^h\ket{1} = \ket{1}\).
With the identity state fixed, we can define the other states as \(\ket{g} \equiv M^+_g\ket{1}\).
Finally, we need to construct a Hamiltonian that projects into the subspace \(\mathcal{H}_G = \vspan\{\ket{g}: g\in G\}\).

In the standard construction described in Section~\ref{sec:general-construction}, the physical system consists of \(\abs{G}\) spin-\(\frac{1}{2}\) degrees of freedom, whose Hilbert space \(\mathcal{H}\) is \(\C^{2\abs{G}}\), with a symmetry group \(\Mon_{\abs{G}}\) generated by spin-flips and permutations.
We can choose a \(\sigma^z\)-computational basis for \(\mathcal{H}\) and combine the values of the \(\sigma^z\) operators as a \(\pm 1\)-vector label for these states.
Then the spin-flips and permutations are equivalent to \(\pm 1\)-monomial matrice multiplying the labels.
The standard construction is based on the fact that the group \(M(G)\) can always be embedded in \(\Mon_{\abs{G}}\) as permutation matrices via the left- and right-regular representations.
We have also chosen \(\mathcal{H}_G\) to be the \((\abs{G} - 2)\)-magnetization subspace of the Hilbert space of \(\abs{G}\) spins, generated by the identity state that the label \(v(1) = [-1, +1, \dots, +1]\).

The standard embedding of \(M(G)\) into \(\Mon_{\abs{G}}\) as the regular representations of \(G\) does not tie us into the choice of spins as the underlying physical degrees of freedom.
It only means that the physical degrees of freedom are labeled by a vector with \(\abs{G}\) entries, which can be regarded as elements of \(\C[G]\), and the multiplication operators acts as permutation matrices \(\ell(h)\) and \(r(h)\) on these vectors.
There remains some freedom in the choice of the identity state, as well as the physical meaning of these states.
The constraints come from the relations \eqref{eq:v-l-r-relations}.
When \(v(1)\) is fixed, we can always define \(v(g) = \ell(g) v(1)\), then the relations \(\ell(h)v(g) = v(hg)\) are automatically satisfied.
The second set of relations \(r(h)v(g) = v(gh^{-1})\) are thus equivalent to \(r(h)\ell(g) v(1) = \ell(g) \ell(h^{-1})v(1)\).
Multiplying \(\ell(h)\ell(g^{-1})\) on both sides, we get \(v(1)\) on the right hand side and \(\ell(h) \ell(g^{-1}) r(h) \ell(g) v(1) = \ell(h)r(h)\ell(g^{-1})\ell(g) v(1) = \ell(h)r(h) v(1)\) on the left, where we have swapped \(\ell(g^{-1})\) and \(r(h)\) then canceled \(\ell(g^{-1})\) with \(\ell(g)\) using the group law.
This means that the state label \(v(1)\) satisfies \eqref{eq:v-l-r-relations} if and only if 
\begin{equation}
\ell(h)r(h)v(1) = v(1) \label{eq:state-label-condition-simplified}
\end{equation}
for all \(h\in G\).
Recall that \(v(1)\) is an element of the group algebra \(\C[G]\), so we can express it using the group basis \(v(1) = \sum_{g \in G} v(1)_g g\).
In this form, the left hand side of \eqref{eq:state-label-condition-simplified} can be expressed as \(\sum_{g\in G} v(1)_g h g h^{-1}\).
Since conjugation is an automorphism of the group, changing dummy variable to \(h g h^{-1}\) preserves the sum.
Renaming the dummy variable to \(g\), the sum becomes \(\sum_{g\in G} v(1)_{h^{-1}g h} g\).
Matching this with the sum on the right hand side, \(\sum_{g\in G} v(1)_g g\), we see that
\begin{equation}
v(1)_{h^{-1}g h} = v(1)_g\ ,\text{ for all } h, g \in G\ .
\end{equation}
Thus the coefficients \(v(1)_g\) must be constant on each conjugacy class of \(G\).
We may define vectors \(u_\chi\) whose \(i\)-th entry is \(1\) if the \(i\)-th group element \(g_i\) belongs to the conjugacy class \(\chi\).
The most general form for a state label \(v(1)\) that satisfies \eqref{eq:state-label-condition-simplified} is a linear combination of these basis vectors, \(\sum_\chi a_\chi u_\chi\).
In the \(S_3\) case, there are three conjugacy classes, \([1] = \{1\}\), \([c] = \{c, c^2\}\), and \([s] = \{s, sc, sc^2\}\), so the corresponding basis for \(S_3\) state labels consists of
\begin{equation}
\begin{split}
  u_{[1]} & = [1, 0, 0, 0, 0, 0]\ , \\
  u_{[c]} & = [0, 1, 1, 0, 0, 0]\ , \\
  u_{[s]} & = [0, 0, 0, 1, 1, 1]\ .
\end{split}
\end{equation}
The most general form of \(v(1)\) for an embedding of an \(S_3\) gauge degree of freedom is \(a_{[1]} u_{[1]} + a_{[c]} u_{[c]} + a_{[s]} u_{[s]} = [a_{[1]}, a_{[c]}, a_{[c]}, a_{[s]}, a_{[s]}, a_{[s]}]\).
Such freedom in the value of \(v(1)\) introduces a great deal of flexibility in possible physical realizations.
Now the entries of \(v(g)\) can be any natural quantum number, not just spins.
For example, we can embed a \(\abs{G}\)-gauge variable into the charge-\(1\) subspace of a system of \(\abs{G}\) electric charges, and let vector labels \(v(g)\) denote the occupation number of the \(\abs{G}\) states.
In that case, we would have \(v(1) = [1, 0, \dots, 0]\).

It is also worth noting that the choice of the identity vector \(v(1)\) can be made independently for the representation of the gauge variables and for the entries of the CGS coupling matrix.
Recall that the coupling matrix is generated from an initial set of couplings \(\begin{bmatrix}\trans{v(1)} &  \dots & \trans{v(1)}\end{bmatrix}\).
If instead of \(v(1)\), we put \(w_j(1) = \sum_{[g]} a_{[g], j} u_{[g]}\) in the \(j\)-th entry, where \(w_j(1)\) is a \(\abs{G}\)-vector whose entries \(a_{[g]}\) are constant within each conjugacy class \([g]\), all of the arguments in \ref{sec:general-construction} still apply, and \(W\) would retain its combinatorial symmetry. 
This means that there are in fact \(m K\) tuning parameters in the coupling matrix \(W\), where \(K\) is the number of conjugacy classes of \(G\).
When we consider this larger parameter space, the flux-fixing term can be absorbed into the choice of the generating entries of \(W\).
This is because the flux-fixing term is proportional to \(\sum_a \mu_a^z\), which is equivalent to shifting \(v(1)\) by a constant vector.

\section{Additional examples and construction for dihedral group \texorpdfstring{\(D_8\)}{D8}}
\label{app:additional-examples}

Under the constraints laid out in Appendix~\ref{app:gauge-embedding-general-requirements}, some gauge groups admit embeddings into fewer physical spins than required by the standard procedure.
This is the case for the quaternion model developed by \citet{Green2023}.
The same is true for the dihedral group \(D_8\), which can be represented as \(4\)-spinors.
Interestingly, the representation uses the same set of spinor states as the \(4\)-spinor representation of the quaternion group \(Q_8\), but the multiplication operators and the resulting coupling matrix will be different.
The smallest embedding of an \(S_3\) gauge variable requires six spins, because no smaller monomial group has an order that is divisible by \(\abs{M(G)} = 36\).
Thus the standard embedding described in Sec.~\ref{sec:s3-gauge-symmetry} is minimal.
Note that in principle the units of the underlying physical system need not be spins, and when more general physical degrees of freedom are involved, for example clock variables, the embedding may require fewer of them.
For example, the symmetry group of \(n\) \(k\)-level clock variables is the generalized monomial group \(\Mon_n(\Z_k)\), which has a richer structure than the \(\{0, 1\}\)-monomial group \(\Mon_n\).
Thus if we have access to \(\Z_3\)-clock variables, the \(S_3\) gauge variable can be realized using only \(4\) of them.

Next we describe a \(4\)-spin embedding of \(D_8\) gauge variables.
We take \(D_8\), the dihedral group of a square to be generated by \(a\), the \(\pi/2\) rotation, and \(x\), the reflection with respect to the \(x\)-axis.
Represented as a matrix subgroup of \(O(2)\), its elements are
\begin{equation}
\begin{aligned}
  e & = \begin{bmatrix}
    1 & 0 \\
    0 & 1
  \end{bmatrix}\ ,&
  a & = \begin{bmatrix}
    0 & -1 \\
    1 & 0
  \end{bmatrix}\ ,\\
  a^2 & = \begin{bmatrix}
    -1 & 0 \\
    0 & -1
  \end{bmatrix}\ ,&
  a^3 & = \begin{bmatrix}
    0 & 1 \\
    -1 & 0
  \end{bmatrix}\ ,\\
  x & = \begin{bmatrix}
    1 & 0 \\
    0 & -1
  \end{bmatrix}\ ,&
  a x & = \begin{bmatrix}
    0 & -1 \\
    -1 & 0
  \end{bmatrix}\ ,\\
  a^2 x & = \begin{bmatrix}
    -1 & 0 \\
    0 & 1
  \end{bmatrix}\ ,&
  a^3 x & = \begin{bmatrix}
    0 & 1 \\
    1 & 0
  \end{bmatrix}\ .
\end{aligned}
\label{eq:d8-elements}
\end{equation}
These can be represented by the following values of a \(4\)-spinor,
\begin{equation}
\begin{aligned}
  v(1) & = \trans{[+1, +1, +1, +1]} & v(x) = \trans{[-1, -1, +1, +1]} \\
  v(a) & = \trans{[-1, +1, +1, -1]} & v(a x) = \trans{[-1, +1, -1, +1]} \\
  v(a^2) & = \trans{[-1, -1, -1, -1]} & v(a^2 x) = \trans{[+1, +1, -1, -1]} \\
  v(a^3) & = \trans{[+1, -1, -1, +1]} & v(a^3 x) = \trans{[+1, -1, +1, -1]}
\end{aligned}
\label{eq:d8-spinor-values}
\end{equation}
The representation of the left multiplication operators are 
\begin{equation}
\begin{aligned}
\ell(e) & = \begin{bmatrix}
  1 & 0 & 0 & 0 \\
  0 & 1 & 0 & 0 \\
  0 & 0 & 1 & 0 \\
  0 & 0 & 0 & 1
\end{bmatrix}
& 
\ell(x) & = \begin{bmatrix}
  -1 & 0 & 0 & 0 \\
  0 & -1 & 0 & 0 \\
  0 & 0 & 1 & 0 \\
  0 & 0 & 0 & 1
\end{bmatrix}
\\
\ell(a) & = \begin{bmatrix}
  0 & 0 & -1 & 0 \\
  0 & 0 & 0 & 1 \\
  1 & 0 & 0 & 0 \\
  0 & -1 & 0 & 0
\end{bmatrix}
& 
\ell(a x) & = \begin{bmatrix}
  0 & 0 & -1 & 0 \\
  0 & 0 & 0 & 1 \\
  -1 & 0 & 0 & 0 \\
  0 & 1 & 0 & 0
\end{bmatrix}
\\
\ell(a^2) & = \begin{bmatrix}
  -1 & 0 & 0 & 0 \\
  0 & -1 & 0 & 0 \\
  0 & 0 & -1 & 0 \\
  0 & 0 & 0 & -1
\end{bmatrix}
&
\ell(a^2 x) & = \begin{bmatrix}
  1 & 0 & 0 & 0 \\
  0 & 1 & 0 & 0 \\
  0 & 0 & -1 & 0 \\
  0 & 0 & 0 & -1
\end{bmatrix}
\\
\ell(a^3) & = \begin{bmatrix}
  0 & 0 & 1 & 0 \\
  0 & 0 & 0 & -1 \\
  -1 & 0 & 0 & 0 \\
  0 & 1 & 0 & 0
\end{bmatrix}
& 
\ell(a^3 x) & = \begin{bmatrix}
  0 & 0 & 1 & 0 \\
  0 & 0 & 0 & -1 \\
  1 & 0 & 0 & 0 \\
  0 & -1 & 0 & 0
\end{bmatrix}
\end{aligned}
\label{eq:d8-l-matrices}
\end{equation}
and those of the right-multiplication operators are
\begin{equation}
\begin{aligned}
r(e) & = \begin{bmatrix}
  1 & 0 & 0 & 0 \\
  0 & 1 & 0 & 0 \\
  0 & 0 & 1 & 0 \\
  0 & 0 & 0 & 1
\end{bmatrix}
&
r(x) & = \begin{bmatrix}
  0 & -1 & 0 & 0 \\
  -1 & 0 & 0 & 0 \\
  0 & 0 & 0 & 1 \\
  0 & 0 & 1 & 0
\end{bmatrix}
\\
r(a) & = \begin{bmatrix}
  0 & 1 & 0 & 0 \\
  -1 & 0 & 0 & 0 \\
  0 & 0 & 0 & -1 \\
  0 & 0 & 1 & 0
\end{bmatrix}
&
r(a x) & = \begin{bmatrix}
  -1 & 0 & 0 & 0 \\
  0 & 1 & 0 & 0 \\
  0 & 0 & -1 & 0 \\
  0 & 0 & 0 & 1
\end{bmatrix}
\\
r(a^2) & = \begin{bmatrix}
  -1 & 0 & 0 & 0 \\
  0 & -1 & 0 & 0 \\
  0 & 0 & -1 & 0 \\
  0 & 0 & 0 & -1
\end{bmatrix}
&
r(a^2 x) & = \begin{bmatrix}
  0 & 1 & 0 & 0 \\
  1 & 0 & 0 & 0 \\
  0 & 0 & 0 & -1 \\
  0 & 0 & -1 & 0
\end{bmatrix}
\\
r(a^3) & = \begin{bmatrix}
  0 & -1 & 0 & 0 \\
  1 & 0 & 0 & 0 \\
  0 & 0 & 0 & 1 \\
  0 & 0 & -1 & 0
\end{bmatrix}
&
r(a^3 x) & = \begin{bmatrix}
  1 & 0 & 0 & 0 \\
  0 & -1 & 0 & 0 \\
  0 & 0 & 1 & 0 \\
  0 & 0 & 0 & -1
\end{bmatrix}
\end{aligned}
\label{eq:d8-r-matrices}
\end{equation}
Curiously, the spinor representations of \(D_8\) in \eqref{eq:d8-spinor-values} are the same as those used for \(Q_8\), and the groups of multiplication operators \(M(D_8)\) and \(M(Q_8)\) are isomoprhic.
Concretely, we can express the left- and right-multiplication operators for the generators of \(D_8)\) in terms of elements of \(Q_8\) as \(\ell(a) = r(-k)\), \(\ell(x) = \ell(-k) r(-i)\), \(r(a) = \ell(-k)\), and \(r(x) = \ell(-i)r(k)\).
In the inverse direction we have \(\ell(i) = \ell(a^3) r(x)\), \(\ell(j) = \ell(a^3) r(a^3 x)\), \(r(i) = \ell(a^2 x) r(a^3)\), and \(r(j) = \ell(a x) r(a^3)\).
This does not mean that the \(D_8\) CGS model is the same as the \(Q_8\) one, because the difference in the group structure still leads to differences in the \(W\) matrix.
For example, consider the rows of the \(W\)-matrix that take the form \([\trans{v(1)}, \trans{v(g)}, \trans{v(g)}]\).
This configuration is possible only if \(g^2 = 1\), so there are as many such rows as there are elements of order \(2\).
In \(D_8\) there are five such elements, but in \(Q_8\) there is only one.
Therefore the \(W\) matrices of the two groups, and therefore the CGS models are different.


\bibliography{NonAbelianCGS}
\bibliographystyle{apsrev4-2}  

\end{document}